\documentclass[12pt]{amsart}

\usepackage{amsfonts,amsmath,latexsym,amssymb,verbatim,amsbsy,amsthm}

\usepackage{pstricks}

\theoremstyle{plain}
\newtheorem{THEOREM}{Theorem}[section]

\newtheorem{theorem}[THEOREM]{Theorem}
\newtheorem{corollary}[THEOREM]{Corollary}
\newtheorem{lemma}[THEOREM]{Lemma}
\newtheorem{proposition}[THEOREM]{Proposition}

\theoremstyle{definition}

\newtheorem{definition}[THEOREM]{Definition}

\theoremstyle{remark}

\newtheorem{remark}[THEOREM]{Remark}
\newtheorem{claim}[THEOREM]{Claim}

\newcommand{\thm}[1]{Theorem~\ref{#1}}

\newcommand{\defin}[1]{Definition~\ref{#1}}

\newcommand{\N}{\ensuremath{\mathbb{N}}}   
\newcommand{\Z}{\ensuremath{\mathbb{Z}}}   
\newcommand{\R}{\ensuremath{\mathbb{R}}}   
\newcommand{\C}{\ensuremath{\mathbb{C}}}   
\newcommand{\T}{\ensuremath{\mathbb{T}}}   
\newcommand{\K}{\ensuremath{\mathbb{K}}}

\def \a {\alpha}
\def \b {\beta}
\def \d {\delta}
\def \g {\gamma}
\def \e {\varepsilon}
\def \f {\varphi}
\def \F {{\bf \Phi}}

\def \l {\lambda}

\def \n {\nabla}
\def \s {\sigma}
\def \th {\theta}
\def \Th {\Theta}
\def \t {\tau}
\def \w {\omega}
\def \O {\Omega}

\def \ba {{\bf a}}
\def \bA {{\bf A}}
\def \bB {{\bf B}}
\def \bC {{\bf C}}
\def \bP {{\bf \Pi}}
\def \Mg {\mathsf{M}}
\def \bpi {\boldsymbol{\pi}}
\def \bpinv {\boldsymbol{\pi}^{-1}}

\def \P {{\bf P}}
\def \Id {{\bf Id}}
\def \id {{\bf id}}
\def \bp {\mathrm{{\bf p}}}
\def \bS {{\bf S}}

\def \bK {{\bf K}}
\def \bM {{\bf M}}
\def \bL {{\bf L}}
\def \bG {{\bf G}}
\def \bT {{\bf T}}
\def \bS {{\bf S}}

\def \bX {{\bf X}}
\def \E {\mathcal{E}}

\def \calkin {\mathcal{C}}

\def \torus {\T^n}

\def \eucno {\R^n \backslash \{0\}}
\def \zinno {\Z^n \backslash \{0\}}

\def \lmax {\l_{\mathrm{max}}}
\def \lmin {\l_{\mathrm{min}}}
\def \llmax {\ell_{\mathrm{max}}}
\def \llmin {\ell_{\mathrm{min}}}

\def \Mp {Ma\~{n}e point}
\def \Ms {Ma\~{n}e sequence}

\def \mmaxm  {\mu^m_{\mathrm{max}}}
\def \mmaxk  {\mu^k_{\mathrm{max}}}
\def \mminm {\mu^m_{\mathrm{min}}}
\def \mmax {\mu_{\mathrm{max}}}
\def \mmin {\mu_{\mathrm{min}}}

\def \frbundle {\mathcal{F}}
\def \sobcon {H_\frbundle^m(\T^n)}
\def \encon {L_\frbundle^2(\T^n)}
\def \enmz {{L}^2_0(\T^n)}
\def \endiv {{L}^2_{\mathrm{div}}(\T^n)}

\def \new {\ba_{\text{new}}}

\def \BX {\bB \bX^{m}}
\def \strip {\mathcal{S}}
\def \flowbox {\mathcal{FB}_{\e,N}}

\def \lan {\langle}
\def \ran {\rangle}
\def \p {\partial}
\def \ra {\rightarrow}
\def \ss {\subset}

\newcommand{\der}[2]{(#1 \cdot \nabla) #2}
\newcommand{\ess}[1]{\s_{\mathrm{ess}}(#1)}
\newcommand{\disc}[1]{\s_{\mathrm{disc}}(#1)}
\newcommand{\ress}[1]{r_{\mathrm{ess}}(#1)}
\newcommand{\sym}[1]{\mathcal{S}^{#1}}
\newcommand{\class}[1]{\mathcal{L}^{#1}}
\newcommand{\sob}[1]{H^{#1}(\torus)}
\newcommand{\sobmz}[1]{H^{#1}_0(\torus)}
\newcommand{\sobdiv}[1]{H^{#1}_{\mathrm{div}}(\torus)}

\newcommand{\sqint}[1]{L^{2}(\T^{#1})}

\def \sp {\s_{\text{p}}}

\DeclareMathOperator{\diver}{div} %
\DeclareMathOperator{\curl}{curl} %
\DeclareMathOperator{\re}{Re} %
\DeclareMathOperator{\Rg}{Rg} %
\DeclareMathOperator{\tr}{Tr} %
\DeclareMathOperator{\Ker}{Ker} %
\newcommand{\rest}[2]{#1\raisebox{-0.3ex}{\mbox{$\mid_{#2}$}}}

\begin{document}

\title{The essential spectrum of advective equations}
\author{Roman Shvydkoy}
\address{University of Illinois at Chicago \\
Department of Mathematics, Statistics, and Computer Science\\
Chicago, IL 60607}
\email{shvydkoy@math.uic.edu} %

\thanks{The author thanks Susan Friedlander,  Yuri Latushkin, and Misha Vishik for stimulating discussions.}%
\date{\today}%

\begin{abstract}
The geometric optics stability method is extended to a general class
of linear advective PDE's with pseudodifferential bounded
perturbation. We give a new short proof of Vishik's formula for the
essential spectral radius. We show that every point in the dynamical
spectrum of the corresponding bicharacteristic-amplitude system
contributes a point into the essential spectrum of the PDE. Generic
spectral pictures are obtained in Sobolev spaces of sufficiently
large smoothness. Applications to instability are presented.
\end{abstract}

\keywords{essential spectrum, ideal fluid, shortwave perturbations,
linear cocycle, bicharacteristic-amplitude system, dynamical
spectrum, Mather semigroup, Ma\~{n}e sequence, Ma\~{n}e point,
approximate eigenfunctions, pseudodifferential operator}

\maketitle

\tableofcontents

\section{Introduction}

The subject of this article is rooted in the geometric stability
method for ideal fluids developed in the early 90's by Friedlander
and Vishik \cite{FV91b,FV93a}, and independently by Lifschitz and
Hameri \cite{Lif91b,LH1}. Studying localized shortwave instabilities
of a general steady flow $u_0$ one is naturally led to consider
solutions of the linearized Euler equation in the WKB form
\begin{equation}\label{WKBintro}
f(x,t) = b(x,t)e^{iS(x,t)/\d} + O(\d)
\end{equation}
where $\d$ is a small parameter. For a very limited class of flows
$u_0$, such a solution may become an exact solution to the (even
nonlinear!) equation  for a finite $\d$. Well-known classical
examples were provided by Craik and Criminale in \cite{CC}, in the
case of a linear vector field $u_0$. For the flow with elliptic
streamlines genuinely three-dimensional perturbations of the form
\eqref{WKBintro} are found to be unstable as shown by numerical
calculations of Pierrehumbert \cite{Pierrehumbert86} and Bayly
\cite{Bayly86}. This elliptic instability is believed to be an
integral part of transition to turbulence in certain laminar flows
\cite{BaylyOrszagHerbert88,Kerswell2002,OrszagPatera80}.

For general equilibrium no explicit solution is available. We
substitute \eqref{WKBintro} into the linearized Euler equation and
set equal the leading order terms on both sides. This gives us
evolution laws for the frequency $\xi = \n S$ and amplitude $b$.
Written in Lagrangian coordinates associated with the basic flow
$u_0$, they form a system of ODE's, called the
bicharacteristic-amplitude system, given by
\begin{subequations}\label{BASintro}
\begin{align}
    x_t &= u(x), \label{BASintro1}\\
    \xi_t &= - \p u^{\top}(x) \xi, \label{BASintro2}\\
    b_t &=-\p u(x) b + 2 \frac{\xi \otimes \xi}{|\xi|^2} \p
    u(x) b.\label{BASintro3}
\end{align}
\end{subequations}
The main result of \cite{FV91b,FV93a,Lif91b,LH1} states that the
exponential growth type of the semigroup generated by the Euler
equation dominates the maximal Lyapunov exponent of the amplitude
equation \eqref{BASintro3}. This provides a sufficient condition for
exponential instability of the general steady flow $u_0$. Papers
\cite{FV92a,Lif94} exhibit a number of examples for which this
condition applies successfully, mainly due to its local nature. In
particular, it is shown that any flow with exponential stretching,
such as a flow with hyperbolic stagnation point, is unstable.

Unlike classical normal modes in a periodic domain, shortwave
perturbations \eqref{WKBintro} are linked to the essential
(continuous) spectrum rather than the point spectrum. It was proved
by Vishik \cite{V96} that for the essential spectral radius
$\ress{\bG_t}$ of the semigroup operator $\bG_t$ the following
formula holds:
\begin{equation}\label{vishikintro}
\ress{\bG_t} = e^{t\mu},
\end{equation}
where $\mu$ is the maximal Lyapunov exponent of the amplitude $b$.
Subsequently, Shvydkoy and Vishik \cite{ShvVish2004} have shown
that, in fact, for any given Lyapunov exponent $\l$ of the amplitude
equation, the circle of radius $e^{t\l}$ contains a point of the
essential spectrum.

The geometric optics method has been applied to many other
non-dissipative equations of ideal hydrodynamics, such as Boussinesq
approximation \cite{FV91b}, SQG \cite{FS2004}, Euler in vorticity
form \cite{LebGod99,Lif94}. Equations with Coriolis forcing were
treated in \cite{LebGod2001,SLJ99}.

The purpose of this present paper is twofold. First, we introduce a
general class of equations, which includes all the equations
mentioned above. In these settings we derive the
bicharacteristic-amplitude system and give a new short proof of
Vishik's formula \eqref{vishikintro}. Second, we give a detailed
description of the essential spectrum in Sobolev spaces.

We consider the following first order linear PDE, which we call an
advective PDE:
\begin{equation}\label{pde0}
    f_t = - \der{u}{f} + \bA f,
\end{equation}
where $u$ is a time-independent smooth vector field and $\bA$  is a
pseudodifferential operator of zero order. We consider
$2\pi$-periodic boundary conditions. For instance, the Euler
equation for incompressible ideal fluid linearized about a steady
state $u$ can be represented in form \eqref{pde0}, where $\bA$ has
principal symbol
\begin{equation}\label{psym}
    \ba_0(x,\xi) = -\p u(x) + 2 \frac{\xi \otimes \xi}{|\xi|^2} \p
    u(x).
\end{equation}
We recognize in \eqref{psym} the right hand side of the amplitude
equation \eqref{BASintro3}. In Section \ref{S:BAS} we show that for
any advective equation the amplitude of a shortwave perturbation
evolves according to the following ODE
\begin{equation}\label{BAS0}
    b_t = \ba_0(x(t),\xi(t)) b,
\end{equation}
where $\ba_0$ is the principal symbol of $\bA$, and $(x(t),\xi(t))$
is the phase flow of \eqref{BASintro1}, \eqref{BASintro2}. We regard
\eqref{BAS0} as a dynamical system over this phase flow. Due to the
Oseledets Multiplicative Ergodic Theorem, we can consider the set of
Lyapunov exponents, of which the maximal one determines the
essential spectral radius over of the semigroup generated by
\eqref{pde0} over $L^2$  via formula \eqref{vishikintro}. This is
proved in \thm{T:ress}. In fact, we prove a version of
\eqref{vishikintro} for any energy-Sobolev space $H^m = W^{2,m}$,
$m\in \R$. In this case the amplitude cocycle is to be augmented by
the frequency, i.e. we consider a new cocycle $|\xi(t)|^m b(t)$,
which we call the $b\xi^m$-cocycle.

Sections \ref{S:description} and \ref{S:sobolev} are devoted to more
detailed description of the essential spectrum.  We show that the
dynamical spectrum (also called Sacker-Sell spectrum) of the
$b\xi^m$-cocycle, in a sense, forms the skeleton of the essential
spectrum. In Theorem \ref{T:sobdesc} we prove the following
inclusions:
\begin{equation}\label{sobinclusionintro}
    \exp\{t \Sigma_{m}\} \ss |\ess{\bG_t}| \ss \exp\{ t [\min \Sigma_m ,
    \max \Sigma_m ] \},
\end{equation}
where $\Sigma_m$ is the dynamical spectrum of the $b\xi^m$-cocycle.
According to a theorem of Sacker and Sell \cite{SSSpec}, the
dynamical spectrum of a $d$-dimensional system is the union of at
most $d$ segments on the real line. So, when dimension of the system
is one, the dynamical spectrum is connected. In this case,
inclusions \eqref{sobinclusionintro} turn into exact identities.
This situation applies, for instance, to all gradient systems of the
form \eqref{pde0}, or to the 2D Euler equation in both vorticity and
velocity form.

The case of large smoothness parameter $m$ is treated in Section
\ref{S:sobolev}. In this case a much more refined description of the
spectrum will be given. If the basic flow $u$ has exponential
stretching of trajectories, the $|\xi|^m$-component of the
$b\xi^m$-cocycle becomes more influential, and eventually takes
control over the spectrum of the whole cocycle. This leads to two
favorable consequences. First, we can control the asymptotics of the
end-points of $\Sigma_m$ with $|m| \ra \infty$, and second, starting
from a certain point $\Sigma_m$ becomes connected due to the fact
that the cocycle $|\xi|^m$ itself is one-dimensional. The precise
quantitative condition on $|m|$ is stated in terms of relevant
Lyapunov exponents in \thm{T:sobolev}. In summary, we will prove the
following result.
\begin{theorem}\label{thmintro}
Suppose $u$ has exponential stretching of trajectories, and let
$|m|$ be large enough. Let us denote $s = \sup_{k \in \R}\{ \min
\Sigma_k \}$ and $S = \inf_{k \in \R}\{ \max \Sigma_k \}$. Then the
following holds:
\begin{itemize}
  \item[1)] $\Sigma_m$ is connected;
  \item[2)] $\min \Sigma_m < s$ and $S < \max \Sigma_m$ ;
  \item[3)] $|\ess{\bG_t}| = \exp\{t \Sigma_{m}\}$;
  \item[4)] $\T \cdot \exp \left\{ t [\min \Sigma_m,s] \cup [S, \max \Sigma_m] \right\}
    \ss \ess{\bG_t}$ ;
\end{itemize}
\end{theorem}
Thus, as we see, $\ess{\bG_t}$ has no circular gaps, and contains
solid outer and inner rings (see Figure \ref{F:semi} for generic
spectral picture).

Parallel results will proved for the spectrum of the generator (the
RHS) of the advective equation \eqref{pde0}. In this case we
restrict the cocycles to the invariant subset $\xi \cdot u(x) = 0$.
We show that under the assumptions of \thm{thmintro} the dynamical
spectrum of the restricted $b\xi^m$-cocycle coincides with the
original spectrum $\Sigma_{m}$. Thus, in addition to the above the
following properties will be proved for the generator $\bL$ (see
\thm{T:sobolevgen}):
\begin{itemize}
\item[5)] $[\min \Sigma_m,s]\cup[S,\max \Sigma_m] + i\R \ss \ess{\bL}$ ;
\item[6)] $\re \ess{\bL} = [\min \Sigma_m,\max \Sigma_m]$.
\end{itemize}
In particular, this implies the Annular Hull Theorem \ref{T:aht},
and other desirable spectral properties.

In the case of the energy space $L^2$, such results are not yet
available. The idea of considering the restricted cocycle has been
already exploited by Latushkin and Vishik \cite{LV2003} in an
attempt to prove the identity between spectral bounds of the
semigroup and generator of the 3D Euler equation. In the 2D case,
however, this result is proved, and as matter of fact, a complete
description of the spectra is given in \cite{SL2003b,SL2003a}. These
are the solid annulus and vertical strip, respectively, for any $m
\neq 0$. The same spectral picture has been found for the SQG
equation in \cite{FS2004}. What these equations have in common is
that their $b$-cocycles have trivial dynamical spectrum $\Sigma_0 =
\{0\}$. In Section \ref{S:sobolev} we show that any advective
equation with trivial dynamical spectrum has annulus--strip
essential spectrum.

Section \ref{s:sss} contains the proof of the above results. Our
main tool is the theory of linear cocycles and Ma\~{n}e sequences.
Some of our statements are novel and have certain applications to
the spectral theory of Mather semigroups. We present these results
in a separate paper \cite{ShvCocycle} that will be published
elsewhere.

\section{Formulation}\label{S:form}

Let $u(x)$ be a smooth vector field on the $n$-dimensional torus
$\T^n$.  Incompressibility of $u$ will be our standing hypothesis,
although it is not always necessary.

We study linear partial differential equations of the form
\begin{equation}\label{pde}
    f_t = - \der{u}{f} + \bA f, \quad t \geq 0
\end{equation}
subject to periodic boundary conditions
$$
f(x+2\pi e_i,t) = f(x,t), \quad i = 1, \ldots ,n,
$$
where $\{e_i\}_{i=1}^n$ are the vectors of the standard unit basis.
A solution $f(x,t)$ assumes values in $\C^d$, and $\bA$ is a
discrete pseudodifferential operator (PDO) defined on smooth
functions by
\begin{equation}\label{pdo}
    \bA f(x) = \sum_{k \in \zinno} \ba(x,k)\hat{f}(k) e^{i k \cdot
    x},
\end{equation}
where $\ba$ is a $d \times d$ matrix-valued symbol
\begin{equation}\label{symbol}
    \ba(x,\xi) : \C^d \ra \C^d,
\end{equation}
defined for all $x \in \T^n$ and $\xi \in  \eucno$.

Throughout the text we impose the following smoothness and growth
assumptions on $\ba(x,\xi)$. The class $\sym{m}$, $m \in \R$,
consists of all infinitely smooth symbols $\ba(x,\xi)$, for $x\in
\torus$ and $\xi \in \eucno$, such that for any multi-indices $\a$
and $\b$ there exists a constant $C_{\a,\b}$ for which the following
estimate holds
$$
|\p_\xi^\a \p_x^\b \ba(x,\xi)| \leq C_{\a,\b} |\xi|^{m - |\a|},
\quad x\in \torus, |\xi| \geq 1.
$$

Unlike in the classical definition of H\"{o}rmander classes
\cite{Hormander3,Shubin} we prefer to allow singularity at $\xi =
0$. As we will see, this is a typical feature of many examples
arising in fluid mechanics. On $\T^n$ (as well as any other compact
manifolds) the singularity can be removed by replacing the original
symbol with a smooth cut-off $\ba(x,\xi)(1- \gamma(\xi))$. This
replacement does not effect the operator $\bA$ so long as
$\gamma(\xi)$ is supported inside the unit ball. We remark that
PDO's of the form \eqref{pdo} with symbols smooth in $\xi$ obey the
same classical principles as in the $\R^n$ case (see \cite{DGY96}).

The class of all pseudodifferential operators of the form
\eqref{pdo} with $\ba \in \sym{m}$ will be denoted by $\class{m}$.

Let $\sob{m}$, $m \in \R$, denote the Sobolev space of $\C^d$-valued
functions on the torus, defined as
\begin{equation}\label{sobolev}
    \sob{m} = \left\{ f: \|f\|_{\sob{m}}^2 = |\hat{f}(0)|^2 + \sum_{k \in \Z^n}
    |k|^{2m}|\hat{f}(k)|^2 < \infty \right\}.
\end{equation}
By the standard boundedness principle  for PDO's
\cite[Theorem~7.1]{Shubin} we have
\begin{equation}\label{boundedness}
\bA: \sob{m} \ra \sob{m-s}
\end{equation}
for any $\bA \in \class{s}$, and any $s,m \in \R$. In fact, in the
case of torus \eqref{boundedness} holds without any smoothness
assumption in the $\xi$-variable, which can be proved by an
application of Minkowski's inequality.

A symbol $\ba \in \sym{0}$ is called {\bf semiclassical} if
\begin{equation}\label{classical}
    \ba = \ba_0 + \ba_1,
\end{equation}
where $\ba_0 \in \sym{0}$ is homogenous of degree $0$ in $\xi$ (i.e.
$\ba_0(x,t\xi) = \ba_0(x,\xi)$), and $\ba_1 \in \sym{-1}$. We call
$\ba_0$ the {\bf principal symbol} of the operator $\bA$. Thus, if
$\ba$ is semiclassical, then
\begin{equation}\label{E:decomp}
\bA =\bA_0 + \bA_1,
\end{equation}
where $\bA_i$ is the PDO with the symbol $\ba_i$. Since $\ba_1 \in
\sym{-1}$, we see from \eqref{boundedness} that $\bA_1$ maps
$\sob{m}$ into $\sob{m+1}$, which embeds back into $\sob{m}$
compactly. Hence, $\bA_1$ is a compact operator on $\sob{m}$.

In this work we consider only semiclassical symbols of class
$\sym{0}$ so that $\bA$ defines a bounded operator on any Sobolev
space, and decomposition \eqref{E:decomp} holds.

Let us denote the right hand side of \eqref{pde} by
\begin{equation}\label{generator}
    \bL f = - \der{u}{f} + \bA f.
\end{equation}
It consists of the advective derivative $- \der{u}{f}$ and bounded
perturbation $\bA f$. The advective derivative generates a
$C_0$-semigroup acting by the rule
$$
f \ra f\circ \varphi_{-t},
$$
where $\f = \{\f_t(x)\}_{t\in\R,\,x\in\T^n}$ is the integral flow of
the field $u(x)$. Hence, $\bL$ itself generates a $C_0$-semigroup
(see Engel and Nagel \cite{Engel-Nagel}). Let us denote it by
$\bG=\{\bG_t\}_{t \geq 0}$. In view of time reversibility of
equation \eqref{pde} the semigroup $\bG$ is invertible, and hence,
it is a group.

\subsection{Constraints}\label{S:constraints} We now introduce a
special class of constraints.

We consider an arbitrary smooth linear bundle $\frbundle$ over the
space of non-zero frequencies $\eucno$. Let us denote its fibers by
$F(\xi) \ss \C^d$, and assume that $F(\xi)$ is $0$-homogenous and
infinitely smooth in the region $\xi \neq 0$. We separately consider
a fiber at zero, $F(0) \ss \C^d$. We call $\frbundle$ {\bf frequency
bundle}.

Given a frequency bundle $\frbundle$, we say that a function $f$ on
$\T^n$ satisfies the {\bf frequency constraints} determined by
$\frbundle$ if $\hat{f}(k) \in F(k)$, for all $k \in \Z^n$.

Let $\bp(\xi): \C^d \ra \C^d$ denote the orthogonal projection onto
$F(\xi)$. According to our assumptions on the fibers $F(\xi)$, $\bp$
is a classical symbol of class $\sym{0}$. For example, the
incompressibility constraint, $\diver{f} = 0$, corresponds to
\begin{align}
F(\xi) & = \{b: b\cdot \xi = 0\}, \label{divfiber}\\
\bp(\xi) & =\id - \frac{\xi \otimes \xi}{|\xi|^2}.\label{divproj}
\end{align}

We introduce the corresponding Sobolev spaces subject to
constraints,
\begin{equation}\label{sobconst}
    \sobcon = \{ f \in H^m(\T^n): \hat{f}(k) \in F(k)\},
\end{equation}
and the orthogonal projection
\begin{gather}
\bP : \sob{m} \ra \sobcon \label{projection1}\\
\widehat{\bP f}(k) = \bp(k) \hat{f}(k). \label{projection2}
\end{gather}
We use special notation $\sobmz{m}$ for the space of mean-zero
functions and $\sobdiv{m}$ for divergence-free fields. If $m=0$, we
write $\encon$, $\enmz$, and $\endiv$.

In the sequel, if constraints are given, we assume that they are
respected by equation \eqref{pde}. In other words, $\bG$ leaves
$\sobcon$ invariant. We note that under this assumption we can still
consider the semigroup $\bG$ on the whole space $\sob{m}$, which
corresponds to solving \eqref{pde} without any constraints.

\subsection{Essential spectrum}\label{ss:esssp} We now briefly state the definition
of essential spectrum used in this paper.

For any closed operator $\bT$ on a Banach space $X$ we use the
following classification of the spectrum (following Browder
\cite{Browder}). A point $z \in \s(\bT)$ is called a point of the
{\bf discrete spectrum} if it satisfies the following conditions:
\begin{itemize}
    \item[(DS1)] $z$ is an isolated point in $\s(\bT)$;
    \item[(DS2)] $z$ has finite multiplicity, i.e. $\bigcup_{r=1}^\infty
    \mathrm{Ker}(z - \bT)^r = N$ is  finite dimensional in $X$;
    \item[(DS3)] The range of $z-\bT$ is closed.
\end{itemize}
Otherwise, $z$ is called a point of the {\bf essential spectrum}.
Thus,
\begin{equation}\label{specsplit}
    \s(\bT) = \ess{\bT} \cup \disc{\bT}.
\end{equation}
We note that if  $\bT$ is bounded, then condition (DS3) follows from
(DS1,DS2).

Let $\ress{\bT}$ denote the radius of $\ess{\bT}$, and let $\calkin$
be the Calkin algebra over $X$. According to Nussbaum \cite{Nuss},
we have
\begin{equation}\label{nuss}
    \ress{\bT} = \lim_{n\ra \infty} \|\bT^n\|_\calkin^{1/n}.
\end{equation}

Concerning spectrum of a semigroup we recall that the discrete part
obeys the spectral mapping property:
\begin{equation}\label{discexp}
    \disc{\bG_t} \backslash \{0\} = e^{t \disc{\bL}}, \ t \geq 0,
\end{equation}
while the essential part may fail to satisfy it. Generally, we only
have the inclusion (see \cite{Engel-Nagel})
\begin{equation}\label{essexp}
e^{t \ess{\bL}} \ss \ess{\bG_t} \backslash \{0\}.
\end{equation}

\section{The bicharacteristic-amplitude system}\label{S:BAS}

\subsection{Derivation}
We now would like to investigate asymptotic behavior of solutions to
\eqref{pde} with initial data given by a highly oscillating wavelet
localized near some point $x_0 \in \T^n$:
\begin{equation}\label{initial}
f_0(x) = b_0(x)e^{i \xi_0 \cdot x /\d}.
\end{equation}
We consider solution in the geometric optics form
\begin{equation}\label{WKB}
f(x,t)=b(x,t)e^{iS(x,t)/\d} + O(\d),
\end{equation}
where $\n_x S(x,t) \neq 0$, for all $x\in\T^n$ and $t\geq 0$. On the
next step we extract evolution laws for the amplitude $b$ and the
phase (eikonal) $S$ by substituting $f(x,t)$ into equation
\eqref{pde}; but first, we need to find an asymptotic formula for
$\bA f$.

\begin{theorem}\label{T:shubin} Suppose $\bA \in \class{0}$ is a pseudodifferential
operator with semiclassical symbol $\ba(x,\xi)$ so that
decomposition \eqref{classical} holds. Let
$$
f_\d(x) = b(x) e^{i S(x)/\d},
$$
where $b,\, S \in C^\infty(\T^n)$ and $\n S(x) \neq 0$ on the
support of $b$. Then the following asymptotic formula holds, as $\d
\ra 0$,
\begin{equation}\label{AWKB}
    \bA f_\d(x) = \ba_0 (x, \n S(x)) f_\d(x) + O(\sqrt{\d}),
\end{equation}
where the constant in the $O$-term depends on $b$ and $S$.

If, specifically, $S(x) = \xi_0 \cdot x$, for some $\xi_0 \in
\eucno$, then $O(\sqrt{\d})$ in formula \eqref{AWKB} can be improved
to $O(\d)$.
\end{theorem}
\begin{proof}
Formula \eqref{AWKB} is  a particular case of \cite[Theorem
18.1]{Shubin} with the parameters taken $m=\d=0$, $N=\rho = 1$ in
the notation of \cite{Shubin}.

For the second part, let us assume for simplicity that the Fourier
transform of $b(x)$ is supported in the ball of radius $R$. Then we
have
\begin{align*}
      \bA f_\d(x) &= \sum_{k \in \zinno} \ba(x,k)\hat{b}(k - \xi_0 \d^{-1})
                  e^{i k \cdot x} \\
              &= e^{i \xi_0 \cdot x / \d} \sum_{|k| \leq R} \ba(x,k + \xi_0
                  \d^{-1})\hat{b}(k) e^{i k \cdot x} .
\end{align*}
According to our assumptions on $\ba$, we obtain
$$
\ba(x, k +\xi_0 \d^{-1}) = \ba_0 (x, \d k + \xi_0) + \ba_1(x,k+
\xi_0 \d^{-1}).
$$
Since $|k|$ is bounded and $|\ba_1(x,\xi)| \leq C |\xi|^{-1}$, we
see that
\begin{align*}
\ba_0(x, \d k + \xi_0) &= \ba_0(x,\xi_0) + O(\d) ,\\
\ba_1(x,k+ \xi_0 \d^{-1}) &= O(\d).
\end{align*}
This implies
\begin{equation}\label{Aexp}
\bA f_\d(x) = \ba_0(x,\xi_0)b(x)e^{i \xi_0 \cdot x / \d}+ O(\d).
\end{equation}\label{pagethis1}
The argument for general $b$ is more technical but similar. \qed
\end{proof}

Now we are in a position to use equation \eqref{pde}. We substitute
$f(x,t)$ into \eqref{pde} using \eqref{AWKB}. Neglecting the terms
that vanish as $\d \ra 0$, and canceling the exponent, we obtain
$$
b_t+ \frac{i}{\d} b S_t = - \der{u}{b} -\frac{i}{\d}b \der{u}{S} +
\ba_0(x,\n S) b.
$$
This yields the following two equations
\begin{subequations}\label{subeq}
\begin{align}
b_t &=- \der{u}{b}+ \ba_0(x,\n S) b \label{subb}\\
S_t &=-\der{u}{S}.\label{subeqS}
\end{align}
\end{subequations}
It follows directly from \eqref{subeqS} that the phase is given by
$$
S(x,t) = \xi_0 \cdot \f_{-t}(x).
$$

We take the gradient of \eqref{subeqS} to obtain, with $\xi(x,t) =
\n S(x,t)$,
\begin{equation}\label{subeqSS}
\xi_t =-\der{u}{\xi} - \p u^{\top} \xi.
\end{equation}
Rewriting equations \eqref{subb} and \eqref{subeqSS} in the
Lagrangian coordinates associated with the flow $x \ra x(t)=
\varphi_t(x)$ we arrive at a {\bf bicharacteristic-amplitude system
(BAS)} of ODE's given by
\begin{subequations}\label{BAS}
\begin{align}
    x_t &= u(x) \label{BASx},\\
    \xi_t &= -\partial u(x)^\top \xi, \label{BASxi}\\
    b_t &= \ba_0(x,\xi)b, \label{BASb}
\end{align}
\end{subequations}
subject to initial conditions $x(0) = x_0 \in \torus$, $\xi(0) =
\xi_0 \in \eucno$, $b(0) = b_0 \in \C^d$, and the constraint $b_0\in
F(\xi_0)$.

\subsection{Preservation of constraints}\label{S:pres}
Let $\frbundle$ be frequency constraints imposed on \eqref{pde}.
Then any solution $f$ has to satisfy $\hat{f}(k) \in F(k)$. For
solutions of the form \eqref{WKB}, in the asymptotic limit $\d \ra
0$, this implies the following condition on the amplitude
\begin{equation}\label{preserveb}
b(t) \in F(\xi(t)), \quad t \geq 0.
\end{equation}
This condition however does not automatically hold for solutions of
\eqref{BASb} even if it holds for initial time $t=0$.

To overcome this deficiency we find a new operator $\bA_\text{new}$
with principal symbol $\new$ which replaces the original $\bA$ in
the definition of $\bL$ such that the action of $\bL$ on functions
from $\sobcon$ is the same (so that the group $\bG$ remains
unchanged on $\sobcon$), while solutions to the new amplitude
equation
\begin{equation}\label{newampl}
b_t = \new(x,\xi) b
\end{equation}
satisfy \eqref{preserveb}.

We naturally make use of the identity $\bL = \bP \bL$, which holds
on functions from $\sobcon$ by the invariance. We notice that $\bL$
is a pseudodifferential operator with the symbol
$$
-i u(x) \cdot \xi + \ba_0(x,\xi) + \ba_1(x,\xi).
$$
Composing $\bL$ with $\bP$, which has symbol $\bp(\xi)$, we obtain
the product of symbols up to $\sym{-1}$
$$
-i u(x) \cdot \xi + \tilde{\ba}_0(x,\xi) + \bp(\xi)\ba_0(x,\xi) +
\tilde{\ba}_1(x,\xi),
$$
where $\tilde{\ba}_1 \in \sym{-1}$, and $\tilde{\ba}_0$ has entries
$$
\tilde{\ba}_{kl} = -\p u^\top(x) \xi \cdot \n \bp_{kl}(\xi), \quad
k,l=1,\ldots,d.
$$

Notice that for any bicharacteristic curve $(x(t),\xi(t))$, which is
a solution of \eqref{BASx}--\eqref{BASxi}, we have
\begin{equation}\label{entry}
\tilde{\ba}_{kl}(x(t),\xi(t)) =  \frac{d}{dt} \bp_{kl}(\xi(t)).
\end{equation}
So, we obtain the identity $\tilde{\ba}_0 =\bp_t$, where the time
derivative in taken along the bicharacteristics. Let us set
\begin{equation}\label{equivsym}
\new(x,\xi) = \bp(\xi)\ba_0(x,\xi) + \bp_t(\xi).
\end{equation}

We claim that if this symbol is used in the formulation of the
amplitude equation, then $b(t) \in F(\xi(t))$ for all $t$ provided
initially $b_0 \in F(\xi_0)$.

Indeed, by \eqref{newampl} and \eqref{equivsym}, we infer
\begin{equation*} \frac{d}{dt} (\id - \bp)b  = b_t - \bp_t b -
\bp b_t = \new b - \bp_t b - \bp \new b = \bp\bp_t b.
\end{equation*}
From the identity $\bp = \bp^2$ it follows that $\bp_t = \bp\bp_t +
\bp_t\bp$. Continuing the previous line we obtain
$$
\frac{d}{dt} (\id - \bp)b = \bp_t (\id - \bp)b,
$$
and the claim follows from Gr\"{o}nwall's Lemma.

It is easy to check that under the incompressibility constraint
given by \eqref{divfiber}--\eqref{divproj},  transformation
\eqref{equivsym} takes the form
\begin{equation}\label{incomptrans}
\new(x,\xi) = \ba_0(x,\xi) + \frac{\xi \otimes \xi}{|\xi|^2} \left( \p
u(x) - \ba_0(x,\xi)\right).
\end{equation}

\begin{remark}\label{remark1}
We conclude this section with a general convention for the rest of
the paper. If constraints are given, we assume that the symbol has
been modified (if necessary) as above so that the BAS preserves the
constraints.
\end{remark}

\subsection{Examples}\label{S:examples}
In this section we provide a list of examples of equation
\eqref{pde}, which arise from linearizing well-known laws of ideal
fluid dynamics. In all examples derivation of the principal symbol
can be carried out using the standard calculus of PDO. We illustrate
it on the Euler equations.

The Euler equations in velocity form, in any spacial dimension $n =
d$, are given by
\begin{subequations}\label{3Dvel}
\begin{gather}
       u_t + \der{u}{u} + \n p = 0, \\
       \diver u = 0.
\end{gather}
\end{subequations}

Let $u(x)$ be a smooth equilibrium solution of \eqref{3Dvel}. The
linearized equation takes the form
\begin{gather*}\label{3Dvellin}
    f_t = - \der{u}{f} - \der{f}{u} - \n p, \\
    \diver f = 0.
\end{gather*}
Let us rewrite it as follows
\begin{equation}\label{3Dvelpde}
    f_t = - \der{u}{f} + \der{f}{u} -2\der{f}{u} - \n p.
\end{equation}
The first two terms form the Lie bracket of $u$ and $f$, which is
divergence-free. Therefore, the Leray projection applies only to the
third term. So, the pseudodifferential operator $\bA$ can be written
as
\begin{multline*}
\bA f(x) = \p u(x) f(x) - 2\sum_{k \in \zinno} \left( \id - \frac{k
\otimes k}{|k|^2}\right) (\p u \,f)\,\hat{}\,(k) e^{i k \cdot
    x} \\
    = \sum_{k \in \zinno} \left( 2 \frac{k \otimes k}{|k|^2} - \id
\right) (\p u \,f)\,\hat{}\,(k) e^{i k \cdot
    x}.
\end{multline*}

We see a composition of two PDOs with symbols $\frac{\xi \otimes
\xi}{|\xi|^2}$ and $\p u(x)$. According to the Composition Formula
\cite{Shubin}, the symbol of the product is equal to the product of
symbols plus a symbol $\ba_1(x,\xi)$ of class $\sym{-1}$. Thus, we
obtain a decomposition $\ba = \ba_0 + \ba_1$ with the principal part
given by
\begin{equation}\label{3Dvelazero}
\ba_0(x,\xi) = \left( 2 \frac{\xi \otimes \xi}{|\xi|^2} - \id
\right)\p u(x).
\end{equation}
It is clear that the equation respects the incompressibility
constraint.

Similarly, we obtain the following examples.

\begin{itemize}
    \item \emph{Simple transport, 2D Euler for vorticities, Charney-Hasegawa-Mima} \cite{Swaters}:
    \begin{equation}
    b_t = 0,\label{BAS:trans}
    \end{equation}

    \item \emph{Euler for velocities} (see
    \cite{FriedLif2003,Kerswell2002} and references
    therein):
     \begin{equation}\label{BAS:eulervel}
      b_t =\left( 2 \frac{\xi \otimes \xi}{|\xi|^2} - \id \right)\p u(x) b.
     \end{equation}

    \item \emph{Euler for velocities with Coriolis forcing} \cite{SLJ99}:
    \begin{equation}\label{BAS:Eulercorvel}
    b_t =\left( 2 \frac{\xi \otimes \xi}{|\xi|^2} - \id \right)\p u(x)
    b + 2\left( \frac{\xi \otimes \xi}{|\xi|^2} - \id \right) \O\times b.
    \end{equation}

    \item \emph{Euler for vorticities} \cite{LebGod99,Lif94}:
     \begin{equation}\label{BAS:Eulervor}
     b_t = \p u(x)b - \frac{\w(x)  \cdot
     \xi }{|\xi|^{2}} \xi \times b.
     \end{equation}

    \item \emph{Euler for vorticities with Coriolis forcing} \cite{LebGod2001}:
     \begin{equation}\label{BAS:Eulervorcor}
     b_t =\p u(x)b - \frac{(\w(x) + 2 \O) \cdot
     \xi }{|\xi|^{2}} \xi \times b.
     \end{equation}

  \item \emph{Boussinesq approximation} \cite{FV91b}:
    \begin{subequations}\label{BAS:Bouss}
    \begin{align}
      b_t &=\left( 2 \frac{\xi \otimes \xi}{|\xi|^2} - \id \right)\p u(x) b +
      r\left( \id - \frac{\xi \otimes \xi}{|\xi|^2} \right) \n \Phi(x), \label{BAS:Bouss1}\\
      r_t &= -b \cdot \n \rho_0(x).\label{BAS:Bouss2}
    \end{align}
    \end{subequations}

 \item \emph{Camassa-Holm (Euler-$\a$)} \cite{FH2004}:
    \begin{equation}\label{BAS:CH}
    b_t = \left( \frac{\xi \otimes \xi}{|\xi|^2} - \id \right)\p
    u^\top (x) b +\frac{\xi \otimes \xi}{|\xi|^2}\p u(x) b.
    \end{equation}

    \item \emph{Non-relativistic superconductivity}:
    \begin{equation}\label{BAS:norelat}
    b_t =\left( 2 \frac{\xi \otimes \xi}{|\xi|^2} - \id \right)\p u(x)
    b + \left( \id - \frac{\xi \otimes \xi}{|\xi|^2} \right) B \times b.
    \end{equation}

    \item \emph{Surface quasi-geostrophic equation } \cite{CMT94,FS2004,Pedlosky}:
    \begin{equation}\label{BAS:SQG}
     b_t  = i \frac{\xi^\perp \cdot \n \th (x)}{|\xi|} b.
    \end{equation}

 \item \emph{Kinematic dynamo} \cite{Arnold-Khesin}:
     \begin{equation}\label{BAS:kinem}
      b_t = \p u(x) b.
     \end{equation}

\end{itemize}

\subsection{BAS as a dynamical system}\label{DynCont}
The first two equations in \eqref{BAS} form a Hamiltonian system on
the symplectic manifold $\O^n = T^*\T^n \backslash \{0\}$ with the
Hamiltonian
$$
H(x,\xi) = u(x) \cdot \xi.
$$
We note that on the torus the cotangent bundle is trivial, i.e.
$T^*\T^n = \T^n \times \R^n$.

The corresponding phase flow defines a Lebesgue-measure preserving
transformation of $\O^n$ given by
\begin{equation}\label{chiflow}
    \chi_t : (x_0,\xi_0) \ra
    \left(\varphi_t(x_0), \p \f_t^{-\top}(x_0)\xi_0 \right),
\end{equation}
where $\p \f_t(x)$ denotes the Jacobian matrix of $\f_t$, and $\p
\f_t^{-\top}(x)$ is its inverse transpose. In terms of this flow the
amplitude equation \eqref{BASb} can be written as
\begin{equation}\label{b}
b_t = \ba_0(\chi_t(x_0,\xi_0))b.
\end{equation}

According to our Remark \ref{remark1}, \eqref{b} defines a dynamical
system on  bundle $\frbundle$ over $\O^n$ with  fibers
$\bpinv(x,\xi)=F(\xi)$. The fundamental solution of \eqref{b}
defines a smooth linear cocycle over the phase flow $\chi$ (see
\cite{SSSpec})
$$
\bB_t(x_0,\xi_0): b_0 \ra b(t,x_0,\xi_0,b_0),
$$
which maps $F(\xi_0)$ into $F(\xi(t))$. We call it $b$-cocycle.

Along with the phase flow $\chi$ we consider its projectivization,
$\overline{\chi}$, onto the compact space $\K^n = \torus \times
\mathbb{S}^{n-1}$, where $\mathbb{S}^{n-1}$ is the unit sphere in
$\R^n$. The map $\overline{\chi}_t$ is defined by the rule
\begin{equation}\label{chiproj}
    \overline{\chi}_t : (x_0,\xi_0) \ra \left(\varphi_t(x_0), \frac{\p \f_t^{-\top}(x_0)\xi_0}{|\p
    \f_t^{-\top}(x_0)\xi_0|}\right).
\end{equation}
Since $\ba_0$ is $0$-homogenous in $\xi$, the amplitude equation
takes the form
\begin{equation}\label{b-proj}
    b_t = \ba_0 (\overline{\chi}_t(x_0,\xi_0))b,
\end{equation}
and hence the $b$-cocycle can be considered over the compact space
$\K^n$ on the projectivized bundle $\frbundle$.

The exponential growth type of $\bB$ is defined by the maximal
Lyapunov exponent
\begin{equation}\label{fluidexp}
\mmax = \lim_{t \ra \infty}  \frac{1}{t} \log \sup_{(x,\xi) \in
\K^n} \|\bB_t(x,\xi)\|,
\end{equation}
where the norm is taken over the fiber $F(\xi)$. It is also equal to
the largest Lyapunov exponent provided by the Multiplicative Ergodic
Theorem for all $\chi$-invariant measures (see \cite[Theorem
8.15]{CL}).

\subsection{Reduction to $L^2$ and the
$b\xi^m$-cocycle}\label{ss:red} In this section we describe a
general procedure that will be used to obtain results concerning
spectrum on $\sobcon$ automatically from the case $m=0$. TO this
end, we introduce another advective equation
$$
f_t = -\der{u}{f} + \bA_m f
$$
with the right hand side $\bL_m$ on $\encon$ being equivalent to the
original $\bL$ on $\sobcon$ via a similarity relation
\begin{equation}\label{similarityrel}
    \bL_m = \bM_m \bL \bM_m^{-1},
\end{equation}
where $\bM_m$ is an isomorphism between $\sobcon$ and $\encon$.

Let $\bM_m$ be the Fourier multiplier with a smooth scalar
non-vanishing symbol equal to $|\xi|^m$ for $|\xi| > 1/2$. Clearly,
$\bM_m : \sobcon \ra \encon$ is an isomorphism. Consider the
operator $\bL_m$ given by \eqref{similarityrel}. By the Composition
Formula for PDO, we have
$$
\bL_m = -\der{u}{} + \bA_m
$$
where $\bA_m$ is a PDO with principal symbol given by
$$
\ba_m(x,\xi) = \ba_0(x,\xi) - m (\p u^{\top}(x) \xi, \xi)
    |\xi|^{-2} \id,
$$
for all $|\xi| \geq 1$. The corresponding BAS with the amplitude
equation
$$
b_t = \ba_m(x,\xi)b
$$
defines a new cocycle, called $b\xi^m$-cocycle, given by
\begin{equation}\label{BXcocycle}
    \left(\BX\right)_t(x,\xi) = \left|\p \f_t^{-\top}(x) \overline{\xi} \right|^m
    \bB_t(x,\xi), \quad (x,\xi) \in \O^n,
\end{equation}
where $\overline{\xi}$ denotes the unit vector $\xi |\xi|^{-1}$. The
$b\xi^m$-cocycle is defined on the same vector bundle $\frbundle$,
and is $0$-homogenous in $\xi$. Thus, everything said about the
$b$-cocycle remains valid for the $b\xi^m$-cocycle too.

We note that the similarity relation \eqref{similarityrel}
establishes equivalence of both discrete and essential parts of the
spectra. This also concerns the spectra of the corresponding groups.
The general procedure will thus be to prove a result in the
$L^2$-space, and deduce the case of arbitrary $m \in \R$ by
replacing the $b$-cocycle with the $b\xi^m$-cocycle. In particular,
we will use the maximal Lyapunov exponent of the $b\xi^m$-cocycle
defined analogously to \eqref{fluidexp},
\begin{equation}\label{fluidexp2}
\mmaxm = \lim_{t \ra \infty}  \frac{1}{t} \log \sup_{(x,\xi)\in
\K^n} \| \left(\BX\right)_t(x,\xi)\|.
\end{equation}

\section{Essential spectral radius}\label{S:ress}
In this section we establish a formula for the radius of essential
spectrum of the semigroup $\bG$ on any constrained (or not) Sobolev
space.

\begin{theorem}\label{T:ress} Let $\bG$ be the $C_0$-group generated on $\sobcon$, $m \in \R$,
by equation \eqref{pde}, in which $u(x)$ is a smooth divergence-free
vector field. Then the essential spectral radius of $\bG_t$ is given
by the formula
\begin{equation}\label{vishik}
\ress{\bG_t} = e^{\mmaxm t} , \quad t \geq 0.
\end{equation}
\end{theorem}

As discussed in Section \ref{ss:red} it suffices to prove the
theorem only in the case $m=0$.

The proof consists of two main parts. First, we describe the
microlocal structure of the evolution operator $\bG_t$. To this end,
let us define the following PDO
\begin{equation}\label{Spdo}
    \bS_t f(x) = \sum_{k \in \zinno} \bB_t(x,k)\hat{f}(k)e^{i k
    \cdot x},
\end{equation}
and let us put
\begin{equation}\label{T}
\bT_t f = \bP[ (\bS_t f) \circ \f_{-t}].
\end{equation}

We will prove the following proposition.
\begin{proposition}\label{comp}
The following decomposition holds for all $t\in\R$
\begin{equation}\label{E:comp}
    \bG_t =\bT_t + \bK_t,
\end{equation}
where $\bK_t$ is a compact operator on $\encon$.
\end{proposition}

Using Proposition \ref{comp} and Nussbaum's formula \eqref{nuss} we
can rewrite \eqref{vishik} in terms of the operator $\bT_t$:
\begin{equation}\label{ress1}
    \ress{\bG_t} = \lim_{n \ra \infty} \|\bG_{nt} \|_{\calkin}^{1/n}
    =\lim_{n \ra \infty} \|\bT_{nt} \|_{\calkin}^{1/n},
\end{equation}
where $\calkin$ is the Calkin algebra over $\encon$. We now estimate
$\|\bT_{t} \|_{\calkin}$ in terms of $L^\infty$-norm of the
principal symbol of $\bS_t$, which is the $b$-cocycle:
\begin{equation}\label{upsilon}
\Upsilon(t) = \sup_{(x,\xi)\in \K^n} \|\bB_t(x,\xi)\|
\end{equation}
where as usual the norm in understood over the constraint fiber
$F(\xi)$. Notice that by \eqref{fluidexp} we have $\mmax = \lim_{t
\ra \infty} \frac{1}{t} \log \Upsilon(t)$. We will prove the
following proposition.

\begin{proposition}\label{L:estm}
There exists a constant $C>0$ such that the following inequalities
hold for all $t \in \R$
\begin{equation}\label{estimate1}
\Upsilon(t) \leq \|\bT_{t} \|_{\calkin} \leq C\Upsilon(t).
\end{equation}
\end{proposition}

Thus, we obtain
$$
\lim_{n \ra \infty} \|\bT_{nt} \|_{\calkin}^{1/n} = \lim_{n \ra
\infty} (\Upsilon(nt))^{1/n} = \left( \lim_{\t \ra \infty}
\Upsilon(\t)^{1/\t} \right)^t = e^{\mmax t}.
$$

Combining this line with \eqref{ress1} finishes the proof of
Theorem~\ref{T:ress}.

Now we prove the above propositions. The proof of Proposition
\ref{comp} uses only basic calculus of PDO (see Shubin \cite{Shubin}
or H\"{o}rmander \cite{Hormander3}). Proposition \ref{L:estm} is a
consequence of the classical result of Seeley \cite{Seeley65} on
isomorphism between the algebra of PDO of order $0$ modulo compact
operators and the algebra of symbols.

\begin{proof}[of Proposition \ref{comp}]

It suffices to consider the case without any constraints. Indeed, if
there are constrains imposed on \eqref{pde}, then by "forgetting"
about them we can extend $\bG$ to all $\sqint{n}$. Then by the
assumption, \eqref{E:comp} holds for the extended group. Applying
the projection $\bP$ and restricting \eqref{E:comp} to $\encon$ we
obtain \eqref{E:comp} in the general case.

So, let us assume $\bP \equiv \Id$. Then the operator $\bT_t$ takes
the form
$$
\bT_t f =(\bS_t f) \circ \f_{-t}.
$$
By  a straightforward computation, we have
\begin{equation}\label{teq}
\frac{d}{dt}\bT_t f(x) = - \der{u(x)}{\bT_t f(x)} + \left(
\frac{d}{dt}\bS_t f \right) (\f_{-t}(x)).
\end{equation}
Using the amplitude equation \eqref{BASb}, we expand the last term
in \eqref{teq} as follows
\begin{multline}\label{sexp}
\left( \frac{d}{dt}\bS_t f \right) (\f_{-t}(x)) = \rest{ \sum_{k \in
\zinno} \ba_0( \f_t(y) , \p \f_t^{-\top}(y) k) \cdot \\
\cdot \bB_t(y,k)\hat{f}(k)e^{i k \cdot y}}{y = \f_{-t}(x)}.
\end{multline}

Our objective now is to compare this expression with $\bA \bT_t f$.
Let us denote $g = \bS_t f$. One has
\begin{equation}\label{ch}
\bA \bT_t f(x) = \bA (g \circ \f_{-t}) = (\bA'g) \circ \f_{-t},
\end{equation}
where $\bA'$ is a PDO with a semiclassical symbol $\ba' \in
\sym{0}$. By the Change of Variables Formula, there is a symbol
$\ba_1' \in \sym{-1}$ such that
$$
\ba'(\f_{-t}(x),\xi) = \ba_0 (x, \p \f_{-t}^{\top}(x)\xi) +
\ba_1'(x,\xi).
$$
Using the identity $\p \f_{-t}^{\top}(x) = \p
\f_{t}^{-\top}(\f_{-t}(x))$, we can rewrite the previous as follows
\begin{equation}\label{ch2}
\ba'(\f_{-t}(x),\xi) = \rest{\ba_0( \f_t(y) , \p \f_t^{-\top}(y)
\xi)}{y = \f_{-t}(x)} + \ba_1'(x,\xi).
\end{equation}
Continuing \eqref{ch}, we obtain
\begin{align}
\bA \bT_t f(x) &=  \sum_{k \in \zinno} \ba'( \f_{-t}(x) ,
k)\hat{g}(k)e^{i k \cdot \f_{-t}(x)} \notag\\
& = \rest{ \sum_{k \in \zinno} \ba_0( \f_t(y) , \p \f_t^{-\top}(y)
k)\hat{g}(k)e^{i k \cdot y}}{y = \f_{-t}(x)}  \label{at} \\
& + \bC_t^{(1)} f(\f_{-t}(x)), \notag
\end{align}
where $\bC_t^{(1)}$ is a PDO of class $\class{-1}$. Evidently, the
symbol of $\bC_t^{(1)}$ is smooth in time.

The principal term in \eqref{at} involves a composition of two PDOs
with symbols $\ba_0( \f_t(y) , \p \f_t^{-\top}(y) \xi)$ and
$\bB_t(y,\xi)$, while the right hand side of \eqref{sexp} involves a
single PDO with the product of the symbols. By the Composition
Formula the difference of \eqref{sexp} and \eqref{at} is a PDO
$\bC_t^{(2)} \in \class{-1}$ composed with the flow map
$\f_{-t}(x)$. It also follows from this argument that the symbol of
$\bC_t^{(2)}$ is smooth in time.

Thus, we have shown that
$$
\left( \frac{d}{dt}\bS_t f \right) \circ \f_{-t} - \bA \bT_t f =
(\bC_t^{(2)} f) \circ \f_{-t}.
$$
Going back to \eqref{teq}, we obtain
$$
\frac{d}{dt}\bT_t f = - \der{u}{\bT_t f} + \bA \bT_t f +
(\bC_t^{(2)} f) \circ \f_{-t} = \bL \bT_t f + (\bC_t^{(2)} f) \circ
\f_{-t} .
$$
By Duhamel's Principle,
\begin{equation}\label{E:exactrepr}
\bT_t f = \bG_t f + \int_0^t \bG_{t-s} [ (\bC_s^{(2)} f) \circ
\f_{-s}] \,ds.
\end{equation}
Let us put
\begin{equation}\label{E:K}
    \bK_t f = -\int_0^t \bG_{t-s} [ (\bC_s^{(2)} f) \circ
\f_{-s}] \,ds.
\end{equation}
Since the family of operators under the integral is strongly
continuous in $s$ and compact, $\bK_t$ is a compact operator (the
proof of this fact can be found in \cite[p.~164]{Engel-Nagel}). \qed
\end{proof}

\begin{proof}[of Proposition \ref{L:estm}]
Let us note that in the constraint-free case we have
$\|\bT_t\|_{\calkin} = \|\bS_t\|_\calkin$. Inequalities
\eqref{estimate1} then follow from the classical result of Seeley
\cite{Seeley65} on isomorphism of the subalgebra of PDO's in
$\calkin$ and the space of $0$-homogenous symbols.

To extend the result to arbitrary frequency constraints we consider
the trivial extension of $\bB_t$ to all of $\C^d$ acting by the rule
$\bB^0_t(x,\xi) b = \bB_t(x,\xi) \bp(\xi) b$. Let us define the
corresponding PDO:
$$
\bS^0_t f  = \sum_{k \in \zinno} \bB^0_t(x,k)\hat{f}(k)e^{i k
    \cdot x} : \sqint{n} \ra \sqint{n}.
$$
According to Seeley, $\|\bS_t^0\|_{\calkin(L^2)} \cong \Upsilon(t)$,
because $\|\bB_t\| = \|\bB_t^0\|$. So, it suffices to show that
$\|\bS_t^0\|_{\calkin(L^2)} = \|\bT_t\|_{\calkin(L_\frbundle^2)}$.

Let us observe that the inequality $\|\bS_t^0\|_{\calkin(L^2)} \geq
\|\bT_t\|_{\calkin(L_\frbundle^2)}$ follows trivially by restriction
and projection. To prove the opposite inequality we claim that the
operator
\begin{equation}\label{trivcomp}
f \ra (\Id - \bP)[(\bS_t^0 f) \circ \f_{-t} ] : \sqint{n} \ra
\sqint{n}
\end{equation} is compact.
\def \ext {\text{ext}}

Indeed, consider the constraint-free extension of $\bG$ to
$\sqint{n}$ as discussed previously. We denote it by $\bG^\ext$. Let
us also denote $\bB_t^\ext$ the corresponding cocycle obtained  as
the fundamental solution of \eqref{BASb} without the constraint
condition $b \in F(\xi)$. By Proposition \ref{comp}, we have
\begin{equation}\label{extcomp}
\bG_t^\ext = \bT_t^\ext + \text{compact}
\end{equation}
on $\sqint{n}$, where
\begin{align*}
\bT_t^\ext & = (\bS_t^\ext f) \circ \f_{-t},\\
\bS^\ext_t f  & = \sum_{k \in \zinno} \bB^\ext_t(x,k)\hat{f}(k)e^{i
k \cdot x}.
\end{align*}
Restricting \eqref{extcomp} to $\encon$ and applying $\Id - \bP$ we
obtain, by invariance,
$$
0 = (\Id - \bP)[ (\bS_t f) \circ \f_{-t} ] + \text{compact} :
L^2_\frbundle \ra L^2.
$$
Since $\bS_t^0$ is the trivial extension of $\bS_t$ it follows that
\eqref{trivcomp} is compact.

Now let us fix any $\e>0$, and find a compact operator $\bK :
L^2_\frbundle \ra L^2_\frbundle$ such that
$$
\|\bT_t + \bK\|_{L^2_\frbundle} \leq
\|\bT_t\|_{\calkin(L^2_\frbundle)} + \e.
$$
Let us extend $\bK$ to all of $L^2$ by $0$ on the complement of
$L^2_\frbundle$, and denote the extension by $\bK^0$. Then
$$
\|\bP[ \bS_t^0(\cdot) \circ \f_{-t}] + \bK^0 \|_{L^2} = \|\bT_t +
\bK\|_{L^2_\frbundle}.
$$
Writing
$$
\bS_t^0(\cdot) \circ \f_{-t} = \bP[ \bS_t^0(\cdot) \circ \f_{-t}] +
(\Id - \bP)[ \bS_t^0(\cdot) \circ \f_{-t}]
$$
and using our claim we conclude
\begin{align*}
\|\bS_t^0\|_{\calkin(L^2)} & = \|\bS_t^0(\cdot) \circ
\f_{-t}\|_{\calkin(L^2)} \leq \| \bP[ \bS_t^0(\cdot) \circ \f_{-t}]
+ \bK^0 \|_{L^2}  \\ & = \| \bT_t + \bK \|_{L^2_\frbundle} \leq
\|\bT_t\|_{\calkin(L^2_\frbundle)} + \e.
\end{align*}

This finishes the proof of Proposition \ref{L:estm}. \qed
\end{proof}

\subsection{Shortwave asymptotics}

We can use the explicit representation of the compact term in
\eqref{E:comp} given by \eqref{E:K} to justify asymptotic formula
\eqref{WKB} for the geometric optics solutions. Taking into account
the constraints, we consider initially
\begin{equation}\label{mode}
f_\d = \bP[b_0 h_0(x) e^{\xi_0 \cdot x/\d}], \quad b_0 \in F(\xi_0),
\; \d \ll 1.
\end{equation}
From \eqref{E:K} we can see that the integral involves a continuous
family of pseudodifferential operators of class $\class{-1}$.
Consequently, by \thm{T:shubin}, $\bK_t f_\d$ decays like $O(\d)$.

Applying \thm{T:shubin} again to the pseudodifferential operator
$\bS_t$ we obtain the asymptotics of $\bT_t f_\d$. Thus, one has the
following formula, as $\d \ra 0$,
\begin{equation}\label{F:asym}
\bG_t f_\d(x) = \bB_t(\f_{-t}(x),\xi_0) b_0 h_0(\f_{-t}(x)) e^{\xi_0
\cdot \f_{-t}(x)/\d} + O(\d),
\end{equation}
where the constant in the $O$-term depends on $t$ and smoothness of
$h_0$.

One is also interested in the size of time interval on which
\eqref{F:asym} holds. Sacrificing $O(\d)$ to a slower term like
$O(\sqrt{\d})$, we can show that \eqref{F:asym} holds for all $t \in
[0, -c \log{\d} ]$ with $c>0$, and $O$ independent of $t$.

Indeed, the $\d$-order term in $\bT_t f_\d$ is bounded by the
supremum of the $\xi$-derivative of the $b$-cocycle. Since the
$b$-cocycle solves the amplitude equation \eqref{BASb},  $\p_\xi
\bB_t$ solves
$$
\frac{d}{dt}\, \p_\xi \bB_t = \p_\xi \ba_0\, \bB_t + \ba_0\, \p_\xi
\bB_t.
$$
Thus, the norm of $\p_\xi \bB_t$ grows at most exponentially.
Similar analysis can be made for the $\d$-order term arising from
$\bK_t$.

So, we obtain
$$
\bG_t f_\d(x) = \bB_t(\f_{-t}(x),\xi_0) b_0 h_0(\f_{-t}(x)) e^{\xi_0
\cdot \f_{-t}(x)/\d} + e^{C t} O(\d),
$$
for some $C>0$, and $O$ independent of time. We can choose $c$ to be
$(2C)^{-1}$.

\subsection{Unbounded domains}\label{ss:unbound}
In the case of unbounded domains, e.g. $\R^n$ or flow channel $\R
\times [-L,L]$, the integral (or mixed) analogue of
pseudodifferential operators has to be used in the formulation of
\eqref{pde}. As we have seen, the proof of Proposition \ref{comp}
uses only basic theorems of pseudodifferential calculus, and those
apply for unbounded domains too. Thus, the asymptotic formula
\eqref{F:asym} remains valid. From it we deduce the lower bound on
the radius:
\begin{equation}\label{ressunb}
    \ress{\bG_t} \geq e^{\mmaxm t}.
\end{equation}

In the case of the open space $\R^n$ the formula \eqref{vishik} was
proved by Vishik in an unpublished version of \cite{V96} under the
assumption of vanishing velocity at infinity. However, in general
the proof of the lower bound breaks down due to non-compactness of
PDO from class $\class{-1}$. In fact, we will show that for the
Euler equation on the 2D flow channel $  \R\times[-1,1]$ formula
\eqref{vishik} fails.

We use recent results of Z.~Lin \cite{Lin2} as our starting point.

Let
$$
u(x,y) = \langle U(y), 0 \rangle, \quad x\in \R, \; y \in [-1,1],
$$
be a steady parallel shear flow with inflectional profile $U(y)$
satisfying the conditions of \cite{Lin2}. It is proved that the
eigenvalue problem for the 2D Euler in vorticity formulation
\begin{equation}\label{eig vort}
    \s f = \bL f = - \der{u}{f} - (\curl^{-1} f \cdot \n) \w
\end{equation}
has exact channel wave solutions
\def \lap {\Delta}
\begin{equation}\label{wave soln}
f = \lap (\psi(y) e^{i \a x} )
\end{equation}
for all $\s \in [0,\s_0)$, where $\s_0>0$. Here $\psi$ is a function
from $H^2([-1,1])$ with $\psi( \pm 1)=0$.

We denote by $X$ the $L^2$-space over the channel with periodic
boundary conditions on the walls and mean zero condition in the
$y$-direction. On this space $\curl^{-1}$ is well-defined, so $\bL$
generates a $C_0$-semigroup.

The normal modes \eqref{wave soln} constructed by Z. Lin have
infinite energy. In order to put them into $X$ we use a truncation
procedure, which replaces the exact identity \eqref{eig vort} by
approximate identities, turning each $\s$ into an approximate
eigenvalue.

Let $f$ be the normal mode \eqref{wave soln} satisfying \eqref{eig
vort}. Let $\g_N$ be a smooth function with $\g_N(x) = 1$ for $|x|
\leq N$, and $\g_N(x) = 0$ for $|x| > N+1$, and
$\g'_N,\g''_N,\g'''_N$ being uniformly bounded functions. Put
\begin{equation}\label{trunc eigen}
f_N = \lap(\psi(y) \g_N(x) e^{i \a x}).
\end{equation}
Then
\begin{equation}\label{expanded}
    f_N(x,y) = f(x,y) \g_N(x) + \psi(y)\g''_N(x) e^{i \a x} + 2i\a \psi(y)\g'_N(x) e^{i \a
    x}.
\end{equation}

Using \eqref{expanded}, one obtains the following identity
$$
\bL f_N = \g_N \bL f + g_N,
$$
where $g_N(x,y)$ is a smooth function supported in $N \leq |x| \leq
N+1$ and uniformly bounded in $N$. On the other hand, from
\eqref{expanded} we see that
$$
f_N = \g_N f + h_N,
$$
where $h_N$ possesses similar properties. Thus, we obtain
$$
\s f_N - \bL f_N = \s h_N - g_N,
$$
and hence,
$$
\| \s f_N - \bL f_N \|\cdot \|f_N\|^{-1} < C  \|f_N\|^{-1}.
$$

It follows from \eqref{expanded} that $\|f_N\| \sim N^{1/2}$. So,
the sequence $\{f_N \|f_N\|^{-1}\}_{N=1}^\infty$ is a sequence of
approximate eigenfunctions for $\s$.

This shows that the unstable essential spectrum of the generator,
and hence, that of the semigroup, is not empty on $X$. On the other
hand, we can  see from \eqref{BAS:trans} that $\mmax = 0$.

\subsection{Applications to instability}\label{S:app}
In this section we indicate several applications of \thm{T:ress} to
instability. We recall that a steady state $u$ is linearly unstable
if the corresponding semigroup $\bG$ is unbounded. A simple
sufficient condition for instability of $u$ follows directly from
Propositions \ref{comp}, \ref{L:estm}, and their Sobolev space
analogues as explained in Section \ref{ss:red}. We have
\begin{equation}\label{instabineq}
\|\bG_t\|_{\sobcon} \geq \|\bG_t\|_\calkin = \|\bT_t\|_\calkin \geq
\sup_{(x,\xi)\in \K^n} \| \left(\BX\right)_t(x,\xi)\|.
\end{equation}
Hence, we obtain the following corollary.
\begin{corollary}\label{C:Lyapunov} The steady state $u \in
C^\infty(\T^n)$ is unstable in  $\sobcon$ if the product
$$|b(t,x_0,\xi_0,b_0)|\cdot |\xi(t,x_0,\xi_0)|^m$$ is unbounded in $t > 0$ for
at least one set of initial data $(x_0,\xi_0)\in \K^n$, $b_0 \in
F(\xi_0)$.
\end{corollary}

An actual unstable mode $f \in \sobcon$ such that $\|\bG_t f\| \ra
\infty$ can be constructed explicitly. We postpone the details of
this construction to a later text.

Another consequence of \eqref{instabineq} is a sufficient condition
for exponential instability in the metric of $\sobcon$, namely,
$\mmaxm >0$. This condition is satisfied, for instance, by any flow
$u$ with exponential stretching of trajectories, provided $|m|$ is
sufficiently large. We will show now that for most important
equations of type \eqref{pde} on divergence-free fields, exponential
stretching in the flow $u$ implies $\mmax>0$, which means
instability already in the energy space. Our proof is based on a
generalization of the conservation law found by Friedlander and
Vishik for the BAS arising from the Euler equation \cite{FV92a}.

According to our convention stated in Remark \ref{remark1}, we
assume that the symbol $\ba_0$ has been transformed by the rule
\eqref{incomptrans}. One can easily see that every such symbol is
invariant with respect to subsequent applications of the
transformation \eqref{incomptrans}. This implies the identity
\begin{equation}\label{relation}
 (\p u(x)\, \xi,\xi) = (\ba_0(x,\xi)\, \xi, \xi).
\end{equation}

We use the following notation
\begin{equation}\label{determ}
    \lan v_1,v_2,\ldots,v_n \ran = \det [v_1,v_2,\ldots,v_n],
\end{equation}
where the determinant is composed of column-vectors $v_i \in \C^n$.

\begin{theorem}\label{T:law} Suppose that the BAS \eqref{BAS} preserves the
incompressibility constraint $b \perp \xi$. Let
$b_1,b_2,\ldots,b_{n-1}$ be any $n-1$ linearly independent solutions
of the amplitude equation over a common initial point $(x_0,\xi_0)$;
and let $\xi$ be the corresponding solution of the frequency
equation. Then the quantity
\begin{equation}\label{consdet}
\lan b_1, \ldots ,b_{n-1},\xi \ran |\xi|^{-2} \exp\left\{ - \int_0^t
\tr \ba_0(x(s),\xi(s)) d s \right\}
\end{equation}
is independent of $t$.
\end{theorem}
\begin{proof} We start by computing the derivative
\begin{align}
\frac{d}{dt}\lan b_1,\ldots,b_{n-1},\xi \ran &=\lan \ba_0
b_1,\ldots,b_{n-1},\xi \ran + \ldots\notag\\
& + \lan b_1,\ldots,\ba_0 b_{n-1},\xi \ran  + \lan b_1,\ldots,
b_{n-1}, - \p u^{\top} \xi \ran. \label{last term}
\end{align}
We can replace the vector $- \p u^{\top} \xi$ in the last
determinant without changing it by any other vector that is equal to
$- \p u^{\top} \xi$ modulo $F(\xi)$. In particular, we can use
\begin{equation}\label{replace}
\p u^{\top} \xi -2 \frac{\xi \otimes \xi}{|\xi|^2} \p u^{\top}\xi.
\end{equation}
Furthermore, we can replace the first term $\p u^{\top} \xi$ in
\eqref{replace} by $\ba_0 \xi$ since their orthogonal projections to
the line spanned by $\xi$ are equal, as relation \eqref{relation}
shows. To the second term in \eqref{replace} we apply the identity
$$
-2 \frac{\xi \otimes \xi }{|\xi|^2} \p u^{\top} \xi = \xi
\frac{d}{dt}(\ln|\xi|^2).
$$
After these changes we have
\begin{align*}
\lan b_1,\ldots, b_{n-1}, - \p u^{\top} \xi \ran & = \lan
b_1,\ldots, b_{n-1}, \ba_0 \xi \ran + \lan b_1,\ldots, b_{n-1}, \xi
\ran \frac{d}{dt}(\ln|\xi|^2).
\end{align*}

Continuing from \eqref{last term} we obtain
\begin{align*}
\frac{d}{dt}\lan b_1,\ldots,b_{n-1},\xi \ran &= (\tr \ba_0 +
\frac{d}{dt}(\ln|\xi|^2)) \lan b_1,\ldots,b_{n-1},\xi \ran.
\end{align*}
The result now follows by integration. \qed
\end{proof}

The traces can be computed directly in all the examples listed in
Section \ref{S:examples} that are subject to the incompressibility
constraint. This yields the following conservation laws.
\begin{itemize}
    \item Euler for velocities (with or without Coriolis forcing),
    Camassa-Holm:
    \begin{equation}\label{ECMcons}
       \lan b_1,\ldots,b_{n-1},\xi \ran \equiv const.
    \end{equation}

    \item 3D Euler for vorticities (with or without Coriolis
    forcing), kinematic dynamo, superconductivity:
     \begin{equation}\label{EKScons}
       \lan b_1,\ldots,b_{n-1},\xi \ran  |\xi|^{-2} \equiv const.
    \end{equation}
\end{itemize}

\begin{theorem}\label{instaben}
The equations listed above generate exponentially unstable
semigroups on $\endiv$, provided $u(x)$ has exponential stretching
of trajectories.
\end{theorem}
\begin{proof}
Since $\tr \p u^{\top} = 0$, there must exist exponentially growing
and exponentially decaying solutions to the $\xi$-equation
\eqref{BASxi}.

In the case of \eqref{ECMcons} we choose a decaying solution
$\xi(t)$. By the conservation law, there must exist an exponentially
growing solution to the amplitude equation, and hence $\mmax >0$.

In the case of \eqref{EKScons} we choose a growing solution
$\xi(t)$. \qed
\end{proof}

\section{General inclusion theorem}
\label{S:description}

We continue our discussion with more details on the structure of the
essential spectrum. As we see from Theorem \ref{T:ress}, the maximal
Lyapunov exponent of the $b\xi^m$-cocycle contribute a point to the
spectrum. In \cite{ShvVish2004} it was observed that any other
Lyapunov exponent contributes a point in the same way. In this
section we show that, in fact, the entire dynamical spectrum of the
$b\xi^m$-cocycle exponentiates into $|\ess{\bG_t}|$ (Theorem
\ref{T:sobdesc}). Under certain aperiodicity assumption on the basic
flow $\f$ we prove that points from $\Sigma_m$ generate circles in
$\ess{\bG_t}$ (\thm{T:rotinv}).

Similar results will be obtained for the spectrum of the generator
$\bL$ (\thm{T:spgen}). In this case we consider the dynamical
spectrum of the cocycle restricted to the submanifold $u(x)\cdot \xi
= 0$. This condition has already been used in \cite{LV2003} and its
necessity was indicated in \cite{ShvFried-survey}.

First let us briefly recall general definitions and results from the
theory of linear cocycles. Details can be found in \cite{CL,SSSpec}.

\subsection{Cocycles, dynamical spectrum, and \Ms s}\label{ss:cocycle}
Let $\Th$ be a locally compact metric space countable at infinity
(such as $\O^n$ and $\K^n$), and let $\E$ be a finite-dimensional
vector bundle over $\Th$ with projection $\bpi: \E \ra \Th$.  We
consider a continuous flow of homeomorphisms on $\Th$, $\f =
\{\f_t\}_{t \in \R}$, and a linear strongly continuous exponentially
bounded cocycle $\F = \{\F_t(\th)\}_{t \in \R, \th \in \Th}$ over
$\f$ (a linear extension of $\f$).

We say that the cocycle $\F$ is {\bf exponentially dichotomic} if
there exists a continuous projection-valued function $ \P(\th):
\bpinv(\th) \ra \bpinv(\th)$, $\th \in \Th$, and constants $M>0$ and
$\e >0$ such that
\begin{itemize}
    \item[1)] $\P(\f_t(\th))\F_t(\th) = \F_t(\th)\P(\th)$;
    \item[2)] $\|\F_t(\th)\P(\th)\| \leq M e^{-\e t}$, $t>0$;
    \item[3)] $\|\F_t(\th)(\Id-\P)(\th)\| \leq M e^{\e t}$, $t<0$.
\end{itemize}

A point $\l \in \R$ is said to belong to the {\bf dynamical}
spectrum of $\F$ if the rescaled cocycle $e^{-\l t} \F_t$ is not
exponentially dichotomic. We denote the dynamical spectrum of $\F$
by $\Sigma_\F$. A well-known theorem of Sacker and Sell
\cite{SSSpec} states that for any cocycle $\F$ over a compact space
$\Th$, its spectrum $\Sigma_\F$ consists of the union of a finite
number of disjoint intervals
\begin{equation}\label{segments}
  \Sigma_\F = [r_1^-,r_1^+] \cup \ldots \cup [r_p^-,r_p^+],
\end{equation}
where the number of intervals $p$ does not exceed  dimension of the
vector bundle $\E$. The end-points $r_1^-$ and $r_p^+$ are,
respectively, the minimal and the maximal Lyapunov exponents of the
cocycle, while all the other Lyapunov exponents (even indexes)
belong to $\Sigma_\F$, \cite{JPS87}.

We now state a characterization of the dynamical spectrum in terms
of so-called \Ms s. This result can be deduced from works
\cite{Anton84,LatSt91}, although it has not been explicitly stated.
We refer the reader to \cite{ShvCocycle} for an alternative
self-contained proof and generalizations to the infinite-dimensional
case. First let us recall the notion of a \Ms\ introduced in
\cite{LatSchn99} (see also \cite{CL}).

\begin{definition}\label{mane} A sequence of pairs $\{(\th_n,v_n)\}_{n=1}^\infty$, where $\th_n \in \Th$
and $v_n \in \bpinv(\th_n)$, is called a {\bf Ma\~{n}e sequence} of
the cocycle $\F$ if $\{v_n\}_{n=1}^\infty$ is bounded and there are
constants $C>0$ and $c>0$ such that
\begin{subequations}\label{e:mane1}
\begin{align}
|\F_n(\th_n)v_n| &>c, \label{e:mane1a}\\
|\F_{t}(\th_n)v_n| &<C, \text{ for all } 0\leq t\leq 2n,
\label{e:mane1b}
\end{align}
\end{subequations}
for all $n \in \N$.
\end{definition}

\begin{theorem}\label{T:sp}
For any cocycle $\F$ the following are equivalent:
\begin{itemize}
    \item[(i)] $\l \in \Sigma_\F$;
    \item[(ii)] There is a Ma\~{n}e sequence either for the cocycle
$\{e^{-\l t} \F_t\}_{t\in\R}$ or its dual.
\end{itemize}
\end{theorem}

Here by dual cocycle we understand the cocycle over the inverse flow
$\f_{-t}$ given by $\F^{-*}_{-t}(\th)$, where $-*$ denotes the
inverse of adjoint.

We also recall that if the underlying space $\Th$ is compact,
existence of a \Ms\ is equivalent to existence of a \Mp\ introduced
in \cite{Mane77}. The latter is a point $\th_0 \in \Th$ for which
there exists a (Ma\~{n}e) vector $v_0 \in \bpi^{-1}(\th_0)$ such
that
\begin{equation}\label{d:mp}
    \sup_{t \in \R} | \F_t(\th_0) v_0 | < + \infty.
\end{equation}
We present the proof of this simple fact here as it will be used
later in the text.

Let $\{(\th_n,v_n)\}_{n=1}^\infty$ be a Ma\~{n}e sequence for a
cocycle $\F$. Since $\Th$ is compact, we may assume that
$\f_n(\th_n) \ra \th_0$ and $\F_n(\th_n)v_n \ra v_0$. Then, by
\eqref{e:mane1b},
$$
|\F_t(\th_0)v_0| = \lim_{n\ra \infty}
|\F_t(\f_n(\th_n))\F_n(\th_n)v_n| = \lim_{n\ra \infty}
|\F_{t+n}(\th_n)v_n| \leq C,
$$
for all $t \in \R$.

Conversely, if $\th_0$ is a Ma\~{n}e point with Ma\~{n}e vector
$v_0$, then
$$
\th_n = \f_{-n}(\th_0), \quad v_n = \F_{-n}(\th_0)v_0
$$
defines a Ma\~{n}e sequence.

\subsection{Essential spectrum of the group} We now present results
concerning the essential spectrum of the group $\bG$.

We denote by $\Sigma_m$ the dynamical spectrum of the
$b\xi^m$-cocycle. According to the above $\mmaxm$ is the maximal
element of $\Sigma_m$, while $\mminm$ will denote the minimal
element. If $m=0$ we simply write $\Sigma$, $\mmax$, $\mmin$.

\begin{theorem}\label{T:sobdesc}
Let $\bG$ be the $C_0$-group generated by equation \eqref{pde} on
$\sobcon$, $m \in \R$. Then the following inclusions hold:
\begin{equation}\label{sobinclusion}
    \exp\{t \Sigma_{m}\} \ss |\ess{\bG_t}| \ss \exp\{ t [\mminm ,
    \mmaxm ] \}.
\end{equation}
\end{theorem}

\begin{proof} According to Section \ref{ss:red} we can assume, without loss of
generality, that $m = 0$.

In view of \thm{T:ress},  we have $|\ess{\bG_t}| \leq e^{\mmax t}$.
On the other hand, passing to the inverse operator, we get the
identity
\begin{equation}\label{E:inv}
    \ess{\bG_t} = \ess{\bG_{-t}}^{-1}.
\end{equation}
Notice that $\{\bG_{-t}\}_{t \in \R}$ is the $C_0$-group generated
by $-\bL$. The corresponding  amplitude equation is given by
$$
b_t = - \ba_0(\chi_{-t}(x_0,\xi_0))b.
$$
Its solutions define the inverse $b$-cocycle $\bB_{-t}$, whose
dynamical spectrum is equal to $-\Sigma$. So, the maximal element in
this spectrum is $-\mmin$. Using \thm{T:ress} we arrive at the
formula $\ress{\bG_{-t}} = e^{- \mmin t}$. In view of \eqref{E:inv},
this completes the proof of the right inclusion in
\eqref{sobinclusion}.

Now let $\mu \in \Sigma$. We can assume by rescaling that $\mu = 0$.

Since $0 \in \Sigma$, according to \thm{T:sp} there is a Ma\~{n}e
sequence either for the $b$-cocycle or for its dual. It is easy to
see that the dual $b$-cocycle arises from the dual group $\bG^*$ in
the same manner as the $b$-cocycle arises from $\bG$. Since the
essential spectra of $\bG^*_t$ and $\bG_t$ are complex conjugate to
each other, there is no loss of generality to assume that there
exists a Ma\~{n}e sequence for the $b$-cocycle. Let us denote it by
$\{(x_n,\xi_n),b_n\}_{n=1}^\infty$, where $b_n \in F(\xi_n)$.

We now introduce a two-parameter family of functions. Let
$I_{U_n}(x)$ be a smoothed characteristic function of a small open
neighborhood $U_n$ containing $x_n$. Let us define
\begin{align*}
f_{n,\d}(x) &= b_n |U_n|^{-1/2} I_{U_n}(x) e^{i \xi_n \cdot x /\d},\\
g_{n,\d} &= \bP f_{n,\d}.
\end{align*}

By asymptotic formula \eqref{F:asym}, we obtain for each $n \in \N$
\begin{align*}
\bG_n g_{n,\d} (x) &= \bB_n(\f_{-n}(x), \xi_n)b_n
\frac{I_{U_n}(\f_{-n}(x))}{|U_n|^{1/2}} e^{i \xi_n \cdot \f_{-n}(x)
/\d} + O(\d),
\\
\bG_{2n} g_{n,\d} (x) &= \bB_{2n}(\f_{-2n}(x), \xi_n)b_n
\frac{I_{U_n}(\f_{-2n}(x))}{|U_n|^{1/2}} e^{i \xi_n \cdot
\f_{-2n}(x) /\d} + O(\d).
\end{align*}
In view of these identities we can choosing $U_n$ sufficiently small
so that
\begin{equation}\label{e:gn}
\|\bG_n g_{n,\d}\| >c/2 \; \text{ and }\; \|\bG_{2n}g_{n,\d} \| <2C, %
\end{equation}
for every $n \in \N$ and $\d < \d_n$, where $c$ and $C$ are as in
\defin{mane}.

Let us show now that $1 \in |\ess{\bG_t}|$. Indeed, suppose this is
not true. Then $L^2$ admits splitting $L^2 = X_s \oplus X_c \oplus
X_u$ into spectral subspaces corresponding to the part the spectrum
inside, on, and outside the unit ball, respectively. In addition,
$X_c$ is finite-dimensional. Let
$$
g_{n,\d} = g^s_{n,\d} + g^c_{n,\d}+ g^u_{n,\d}
$$
be the corresponding decomposition of $g_{n,\d}$. Since $g_{n,\d}
\ra 0$ weakly as $\d \ra 0$ for each fixed $n$, we obtain $\lim_{\d
\ra 0}\|g^c_{n,\d}\|  =  0$. Hence, by \eqref{e:gn} for sufficiently
small $\d$, we have
\begin{equation}\label{e:gn2}
\|\bG_n g'_{n,\d}\| >c/2 \; \text{ and }\; \|\bG_{2n}g'_{n,\d} \| <2C, %
\end{equation}
where $g'_{n,\d} =g^s_{n,\d}+ g^u_{n,\d}$. Let us fix a $\d$ for
each $n$ so that inequalities \eqref{e:gn2} hold.

Since the exponential type of $\bG_{-t}$ on $X_u$ is negative, there
are $\e>0$ and $M>0$ such that $\|\bG_t g\| \geq M e^{\e t}\|g\|$ on
$X_u$. This implies
$$
\|\bG_{2n}g'_{n,\d}\| \geq C \|\bG_{2n}g^u_{n,\d}\| \geq C M e^{\e
n} \|\bG_n g^u_{n,\d}\|.
$$
Using \eqref{e:gn2} it follows that $\|\bG_n g^u_{n,\d}\| \leq C_1
e^{-\e n}$. Then
$$
\|\bG_n g'_{n,\d}\| \leq C_2 (\|\bG_n g^s_{n,\d}\|+ \|\bG_n
g^u_{n,\d}\|) \leq C_3 e^{- \e n}.
$$
This contradicts \eqref{e:gn2} and hence finishes the proof of the
first inclusion in \eqref{sobinclusion}. \qed
\end{proof}

So far we have imposed no assumption on the basic flow $\f$.
However, the presence of long orbits in $\f$ entails the property of
rotational invariance of the spectrum, as we will see from our next
result. Similar result for the Mather evolution semigroup generated
by the cocycle is well-known (see \cite{CL} and references therein).
In spite of representation \eqref{E:comp}, which implies certain
resemblance with the Mather semigroup, such a result for the
infinite-dimensional group $\bG$ is not immediate and requires a
separate argument.

In the statement of the theorem we use the concept of a \Mp. As we
noted in Section \ref{ss:cocycle} in the case of a compact space \Mp
s and \Ms s can be used interchangeably. Thus, for the
$b\xi^m$-cocycle considered over $\K^n$ both become available.

\begin{theorem}\label{T:rotinv} Let $\bG$ be the $C_0$-group generated by equation
\eqref{pde} on $\sobcon$, and let $\mu \in \Sigma_m$. Suppose that
there exists a point $(x_0,\xi_0)\in \O^n$ such that
\begin{itemize}
    \item[(i)] $(x_0,\xi_0)$ is a \Mp\ for the rescaled $b\xi^m$-cocycle $e^{-\mu t}\BX_t$, or its
    dual;
    \item[(ii)] for every given $N>0$ any open neighborhood of $x_0$ intersects
    an $\f$-orbit of period greater than $N$.
\end{itemize}
Then the following inclusion holds:
\begin{equation}\label{E:rotinv}
    \T \cdot e^{\mu t} \ss \ess{\bG_t}.
\end{equation}
\end{theorem}

If every point of the torus satisfies hypothesis (ii), then the flow
$\f$ is called {\bf aperiodic}. Thus, from Theorems \ref{T:sobdesc}
and \ref{T:rotinv} we obtain the following corollary.

\begin{corollary}\label{c:aperiodic}If the flow $\f$ is aperiodic, then
the following inclusion holds:
$$
\T \cdot e^{t \Sigma_m} \ss \ess{\bG_t}.
$$
\end{corollary}

\begin{proof}[of \thm{T:rotinv}] In view of the reduction argument presented in
Section \ref{ss:red} we may assume that $m=0$.

For definiteness, we assume that there exists a \Mp\ $(x_0,\xi_0)$
and Ma\~{n}e vector $b_0 \in F(\xi_0)$ corresponding to the
$b$-cocycle. In the dual situation we carry out the argument for the
adjoint semigroup using the correspondence between $\bG^*$ and the
dual of $\bB$ pointed out in the proof of \thm{T:sobdesc}.

Also, by rescaling, we can assume that $\mu = 0$.

We proceed in several steps.

{\sc Step 1.} \emph{Construction of approximate eigenfunctions.}

We construct a sequence of approximate eigenfunctions for $\bG_t$ in
the geometric optics form
\begin{equation}\label{envelope}
    f(x) = b(x) e^{iS(x) /\d} + O(\d), \quad \d \ll 1,
\end{equation}
with the amplitude $b(x)$ supported in a flow-box stretched along an
orbit of large period passing near $x_0$. We let both $b$ and $S$
propagate along this orbit according to their respective evolution
laws defined by the BAS. Several preliminary geometric conditions
will have to be settled in order to properly carry out the
construction.

First, by the assumption on $x_0$, for any given $N>0$ we can choose
a point $x_N\in \T^n$ in a vicinity of $x_0$ so that the period of
$x_N$ is greater than $2N+2$. Second, by taking a small perturbation
of $\xi_0$, if necessary, we can replace $\xi_0$ by a $\xi_N$ which
is not orthogonal to $u(x_N)$. In addition, we choose $(x_N,\xi_N)$
so close to the \Mp\ $(x_0,\xi_0)$ that
\begin{equation}\label{apermane}
    \sup_{-N \leq t \leq N} |\bB_t(x_N,\xi_N)b_0| \leq C_1,
\end{equation}
holds for some $C_1>0$ independent of $N$ (see \eqref{d:mp}).

These geometric conditions on $(x_N,\xi_N)$ enable us to define a
flow-box around the orbit through $x_N$ as follows.

For $\e>0$ consider an $(n-1)$-dimensional planar tile $\Xi$
perpendicular to $\xi_N$, and of spacial measurements $\e \times
\ldots \times \e$. Choosing $\e$ small enough, every point $x$ in
the set $\flowbox = \{\f_t(\Xi)\}_{-N \leq t \leq N}$ is uniquely
determined by a $\s \in \Xi$ and $t \in [-N,N]$ so that $x=
\f_t(\s)$. This set is the desired flow-box.

For every $\a \in [0,2\pi]$ we will now construct a sequence of
functions $f_{\d,\e,N}$ in the form \eqref{envelope} such that, for
some $C_2>0$,
\begin{equation}\label{limit}
    \limsup_{\e \ra 0}\ \limsup_{\d \ra 0} \frac{\| \bG_1 f_{\d,\e,N} -
    e^{i\a}f_{\d,\e,N}\|}{\|f_{\d,\e,N}\|} \leq \frac{C_2}{N}.
\end{equation}
Clearly, this is sufficient for proving the lemma.

For every $x \in \flowbox$, $x = \f_t(\s)$, and $\d>0$ we define the
amplitude $b(x)$ as follows
\begin{equation}\label{amplitude}
    b(x) = \b(\s) \g(t) \bB_t(\s,\xi_N)b_N,
\end{equation}
where $\b$ is any function on $\Xi$ of unit $L^2(\Xi)$-norm, and
where $\g$ is a slowly varying tent-shaped function defined as
$\g(t) = 1-|t|N^{-1}$, for $-N \leq t\leq N$, and $\g(t)=0$
otherwise.

Let us define a phase by the rule
\begin{equation}\label{phase}
    \rest{S(x)}{x=\f_t(\s)} = t.
\end{equation}
Observe that $\rest{\n S}{\Xi}$ is proportional to $\xi_N$, and
$S(\f_t(x)) - S(x) = t$ for all $x$ in the flow-box. Taking the
gradient at $x=\s$ we obtain
\begin{align*}
\p \f_t^\top(\s) \n S(\f_t(\s)) &=\n S(\s)\\
\n S(\f_t(\s)) &=\p \f_t^{-\top}(\s)\n S(\s).
\end{align*}
So, up to a constant multiple,
\begin{equation}\label{phasegrad}
    \rest{S(x)}{x=\f_t(\s)} = \p \f_t^{-\top}(\s) \xi_N,
\end{equation}
for all $x\in \flowbox$. Notice that $\n S(x) \neq 0$ on the
flow-box.

Now, we put
\begin{equation}\label{eigenf}
f_{\d,\e,N} = \bP[b e^{iS/\d}].
\end{equation}
 By
\thm{T:shubin}, we conclude that $f_{\d,\e,N}$ is of the form
\eqref{envelope}.

{\sc Step 2.} \emph{Shortwave asymptotics.}

Let us apply $\bG_1$ to $f= f_{\d,\e,N}$. The action of $\bG_1$ on
$f$ with fixed $\e$ and $N$, in the asymptotic limit $\d \ra 0$, is
easily found using routine application of Proposition \ref{comp},
the Change of Variables Formula for pseudodifferential operators,
and \thm{T:shubin} applied to the operator $\bS_t$ defined by
\eqref{Spdo}. As a result, one obtains the following formula
$$
\bG_1 f (x) = \bB_1 (\f_{-1}(x), \n S(\f_{-1}(x))) f(\f_{-1}(x)) +
o(1),
$$
as $\d \ra 0$. So, if $x = \f_t(\s)$, then by \eqref{phasegrad} we
obtain, up to the leading order term,
\begin{align*}
\bG_1 f(x) &= \b(\s) \g(t-1) \bB_1(\f_{t-1}(\s),\p
\f_{t-1}^{-\top}(\s) \xi_N) \bB_{t-1}(\s,\xi_N) b_0 e^{i (t-1)/\d}\\
& = e^{-i/\d} \b(\s) \g(t-1) \bB_{t}(\s,\xi_N) b_0 e^{i t/\d}\\
& = e^{-i/\d} f(x) +e^{-i/\d} \b(\s) (\g(t-1)-\g(t))
\bB_{t}(\s,\xi_N) b_0 e^{i t/\d}.
\end{align*}

Let us take $\d$ of the form $(2\pi k - \a)^{-1}$, $k\in \N$. Then
from the above we conclude
$$
\bG_1 f - e^{i \a} f =  e^{i \a} \b(\s) (\g(t-1)-\g(t))
\bB_{t}(\s,\xi_N) b_0 e^{i t/\d} + o(1).
$$
It is readily seen that the $\limsup$ of the energy norm of the left
hand side, as $\d \ra 0$, is bounded by the energy norm of
$$
\b(\s) (\g(t-1)-\g(t)) \bB_{t}(\s,\xi_N) b_0.
$$

{\sc Step 3.} \emph{Change of variables over the flow-box.}

When performing integration over the flow-box, it is convenient to
switch from $x$- to $(\s,t)$-variables.

To this end, we define a map from $\strip = [-N,N] \times \Xi$ onto
$\flowbox$ by
$$
H(t,\s) = \f_t(\s).
$$
A direct computation shows that
$$
\p H(t,\s) = \left[  u\circ \f_t(\s),\,  \p_{\s_1} \f_t(\s) , \ldots
,\,  \p_{\s_{n-1}} \f_t(\s) \right],
$$
where $\s = (\s_1,\ldots,\s_{n-1})$ is a system of rectangular
coordinates on $\Xi$. Let $e_k$ be the unit vector in the $\s_k$-th
direction. Then
$$
\p H(t,\s) =\p \f_t(\s) \left[ u(\s),\,  e_1  , \ldots ,\, e_{n-1}
\right].
$$
Consequently, the quantity
$$
\kappa(\s)=\left| \det \p H(t,\s) \right| = \left| \det \left[
u(\s),\, e_1 , \ldots ,\,  e_{n-1} \right] \right|,
$$
is independent of $t$. Besides, $\kappa(0) \neq 0$ due to our
assumption that $\xi_N \not \perp u(x_N)$.

{\sc Step 4.} \emph{Proof of \eqref{limit}.} We obtain
$$
\limsup_{\d \ra 0} \frac{\|\bG_1 f - e^{i \a} f\|^2}{\|f\|^2} \leq
\frac{I_1}{I_2},
$$
where in the $(\s,t)$-coordinates,
\begin{align*}
I_1 & = \int_\strip \kappa(\s) \b^2(\s) (\g(t-1)-\g(t))^2
|\bB_{t}(\s,\xi_N) b_0|^2 d\s dt, \\
I_2 & = \int_\strip \kappa(\s) \b^2(\s) \g^2(t) |\bB_{t}(\s,\xi_N)
b_0|^2 d\s dt.
\end{align*}

Now, let us shrink the tile $\Xi$ to the point $x_N$ -- i.e. let $\e
\ra 0$. Then $\b(\s)$ serves as an approximative kernel. We obtain
\begin{align*}
I_1 & \ra \kappa(0) \int_{-N}^{N}(\g(t-1)-\g(t))^2
|\bB_{t}(x_N,\xi_N) b_0|^2 dt, \\
I_2 & \ra \kappa(0) \int_{-N}^{N} \g^2(t) |\bB_{t}(x_N,\xi_N) b_0|^2
dt.
\end{align*}
Notice that $|\g(t-1)-\g(t)| \leq N^{-1}$ by construction. Thus,
using the previous identities and \eqref{apermane}, we estimate
$$
\limsup_{\e \ra 0} \frac{I_1}{I_2} \lesssim
\frac{N^{-2}\int_{-N}^{N}|\bB_{t}(x_N,\xi_N) b_0|^2 dt}{|b_0|^2}
\leq \frac{C_2}{N}.
$$

This finishes the proof of the theorem.\qed
\end{proof}

\subsection{Essential spectrum of the generator} For general advective
equations \eqref{pde} the spectral mapping theorem is unknown.
However, we can  obtain similar results about the essential spectrum
of $\bL$ considering the dynamical spectrum of the $b\xi^m$-cocycle
restricted to the $\chi$-invariant submanifold
$$
\O^n_0 = \{(x,\xi) \in \O^n : u(x) \cdot \xi = 0\},
$$
which is also $0$-homogenous in $\xi$, and projects onto a
submanifold of $\K^n$. Let us denote the spectrum of the cocycle
restricted to $\O^n_0$ by $\Sigma_m^\perp$.

We show below that the analogues of Theorems \ref{T:sobdesc} and
\ref{T:rotinv} hold for $\bL$ if the full dynamical spectrum is
replaced by $\Sigma_m^\perp$.

\begin{theorem}\label{T:spgen}
Let $\bL$ be the operator defined by equation \eqref{pde} on
$\sobcon$, $m \in \R$. Then the following inclusions hold:
\begin{equation}\label{e:sigmaperpincl}
    \Sigma_m^\perp \ss \re \ess{\bL} \ss [\mminm,\mmaxm].
\end{equation}
Furthermore, if there is a \Mp\ $(x_0,\xi_0)\in \O^n_0$
corresponding to $\mu$  satisfying the assumptions (i) and (ii) of
\thm{T:rotinv}, then
\begin{equation}\label{E:rotinvperp}
    \mu + i\R \ss \ess{\bL}.
\end{equation}
In particular, if the flow $\f$ is aperiodic, then one has
\begin{equation}\label{E:rotinvperpaper}
    \Sigma_m^\perp + i\R \ss \ess{\bL}.
\end{equation}
\end{theorem}

\begin{proof} We first notice that the right inclusion follows
immediately from the general inclusion for essential spectra
\eqref{essexp}, and \thm{T:sobdesc}.

Now let $\mu \in \Sigma_m^\perp$. As before, we may assume that $\mu
= 0$, $m=0$, and that there exists a \Mp\ $(x_0,\xi_0) \in \O_0^n$
for the $b$-cocycle. We consider three cases, which we call
aperiodic, periodic, and the case of stagnation point.

{\sc Case 1: aperiodic.}

Suppose $x_0$ is aperiodic, i.e. assumption (ii) of \thm{T:rotinv}
holds. We then aim at proving \eqref{E:rotinvperp}, which in
particular implies \eqref{e:sigmaperpincl}.

We define
$$
f_{\d,\e,N} = \bP[ b e^{iS/\d}e^{i\a t} ],
$$
where all the ingredients are the same as in the proof of
\thm{T:rotinv} except for the phase function. We define $S$ as
follows.

Let $\Xi$ be a planar $(n-1)$-dimensional tile orthogonal to
$u(x_0)$, containing $x_0$, and having spacial measurements $\e
\times \ldots \times \e$. Since $\xi_0 \perp u(x_0)$, $\xi_0 \in
\Xi$. Let $\tilde{\Xi} \ss \Xi$ be the orthogonal complement of
$\xi_0$ in $\Xi$ containing $x_0$. So, the surface
$\{\f_t(\tilde{\Xi})\}_{-N \leq t \leq N}$ is orthogonal to $\xi_0$
at $x_0$.

In the flow-box, defined by $\flowbox = \{\f_t(\Xi)\}_{-N \leq t
\leq N}$, we introduce the following coordinates
\begin{gather*}
x \ra (\tilde{\s},\t,t), \\
x = \f_t (\s), \quad \s = \tilde{\s} + \t \xi_0.
\end{gather*}
Using these coordinates let us define the phase as follows:
\begin{equation}\label{phaseinlemma}
\rest{S(x)}{x = \f_t(\tilde{\s} + \t \xi_0)} = \t.
\end{equation}
Then $S(x) = S(\f_t(x))$ for all $x \in \flowbox$. This implies that
$\n S(x)$ solves the $\xi$-equation, and by construction,
\begin{equation}\label{e:phaseorth}
    u(x) \cdot \n S(x) = 0.
\end{equation}
Using \eqref{e:phaseorth} and \thm{T:shubin} (notice that $\n S(x)
\neq 0$ in the flow-box!), we obtain, as $\d \ra 0$,
$$
\bL f - i\a f = \b(\s) \g'(t) \bB_t(\s,\xi_0)b_0 e^{i\a t}e^{i \t
/\d} + o(1).
$$
The rest of the proof goes along the lines of \thm{T:rotinv}.

{\sc Case 2: periodic.}

In this case we assume $u(x_0) \neq 0$, and there is a $P>0$ and an
open neighborhood of $x_0$, denoted $U_{x_0}$, such that $p(x) <P$
for all $x \in U_{x_0}$, where $p(x)$ denotes the prime period of
$x$. Let us also define the continuous period function
\begin{equation}\label{D:contper}
    p_c(x) = \lim_{\e \ra 0} \sup_{|x-y|<\e}  p(y) .
\end{equation}

For any small $\e >0$, let $\Xi$ be the planar $(n-1)$-dimensional
tile of spacial measurements $\e \times \ldots \times \e$,
orthogonal to $u(x_0)$. Due to the periodicity assumption, the
flow-box, defined by $\mathcal{FB}_\e = \{\f_t(\Xi)\}_{t\in \R}$,
has the shape of the torus. We define the phase $S(x)$ by
\eqref{phaseinlemma} as before. Since $S(x)$ is flow invariant, it
is well-defined in $\mathcal{FB}_\e$.

Our further argument is based on the following claim.
\begin{claim}\label{claim1}
One has $\p \f_{p_c(x_0)}^{-\top}(x_0)\xi_0 = \xi_0$. So, the flow
$\chi$ is $p_c(x_0)$-periodic at $(x_0,\xi_0)$.
\end{claim}
\begin{proof} Notice that $p_c(x)$ is an integer multiple of $p(x)$,
and $p_c$ is a continuous function where it is finite. We have
$\f_{p_c(x)}(x) = x$ for all $x$ in an open neighborhood of $x_0$.
By the Implicit Function Theorem, $p_c(x)$ is differentiable at
$x_0$ and
$$
\p \f_{p_c(x_0)}(x_0) + u(x_0) \otimes \n p_c(x_0) = \id
$$
Hence,
$$
\p \f_{p_c(x_0)}(x_0) = \id -u(x_0) \otimes \n p_c(x_0).
$$
One has
$$
\p \f^{-\top}_{p_c(x_0)}(x_0) = \p \f^{\top}_{ - p_c(x_0)}(x_0) =
\id +\n p_c(x_0) \otimes u(x_0).
$$
Since $\xi_0 \perp u(x_0)$, then clearly, $\p
\f^{-\top}_{p_c(x_0)}(x_0) \xi_0 = \xi_0$. \qed
\end{proof}

Since $(x_0,\xi_0)$ is a \Mp\ for the $b$-cocycle, there exists a
vector $b_0 \in F(\xi_0)$ such that the sequence
$$
\{\bB_{p_c(x_0)}^k(x_0,\xi_0)b_0\}_{k\in\Z}
$$
is bounded. This implies that there is a $\l(x_0) \in \T$ and
$b(x_0) \in F(\xi_0)$ such that
$$
\bB_{p_c(x_0)}(x_0,\xi_0)b(x_0) = \l(x_0) b(x_0).
$$
By continuity, for every $\s \in \Xi$ there exists a $b(\s) \in F(\n
S(\s))$ and $\l(\s) \in \C$ such that the following holds:
\begin{gather}
    \bB_{p_c(\s)}(\s,\n S(\s)) b(\s) = \l(\s)
    b(\s),\label{e:eigencoc}\\
  |\l(\s) - \l(x_0) | = o(1), \text{ as } \e \ra 0.\label{e:eigval}
\end{gather}

Now we define the function $f_{\d,\e}$ as follows:
$$
f_{\d,\e} = \bP[b e^{iS/\d}],
$$
where the amplitude is given by
$$
b(x) = \b(\s) \l(\s)^{-t/p_c(\s)} \bB_t(\s,\n S(\s)) b(\s), \quad x
\in \mathcal{FB}_\e,
$$
and where by $ \l(\s)^{-t/p_c(\s)}$ we understand the principal
branch of the power function.

This amplitude function $b(x)$ is well-defined in the flow-box only
if the equality
\begin{equation}\label{E:star}
p_c(\s) = p(\s)
\end{equation}
holds for all $\s \in \Xi$. It is easy to check that $p(x)$ is a
lower-semicontinuous function, and, as any such function, it is
continuous on a dense $G_\d$-set. 
Thus, the set $A=\{x:p_c(x) = p(x)\}$ is dense and, evidently,
$\f$-invariant in $\T^n$. Moreover, as we noted above, $p_c(x)$ is
an integer multiple of $p(x)$. So, if $x\in A$ and $0<p(x)<\infty$,
then an open neighborhood of $x$ belongs to $A$.

By virtue of the periodicity assumption, we have $0<p(\s)<P$ for all
$\s \in \Xi$. Hence, $A$ has a non-empty intersection with $\Xi$. In
order to ensure that \eqref{E:star} holds on $\Xi$ it suffices to
restrict $\Xi$ to a smaller tile contained in $A$. In the sequel,
$\Xi$ denotes such a restriction.

Now, since the definition of $f_{\d,\e}$ is validated, we show that
$\{f_{\d,\e}\}$ is a sequence of approximate eigenfunctions
corresponding to the point
$$
z = i \arg \l(x_0) / p_c(x_0).
$$

Indeed, routine computations, based on an application of
\thm{T:shubin} and the fact that the $b$-cocycle solves
\eqref{BASb}, reveal the following asymptotic formula for the action
of $\bL$ on $f_{\d,\e}$ at $x = \f_t(\s)$:
\begin{multline*}
\bL f_{\d,\e}  = -\der{u}{f_{\d,\e}} + \bA f_{\d,\e} \\
= \ln(\l(\s))p^{-1}(\s) \b(\s) \l(\s)^{-t/p_c(\s)} \bB_t(\s,\n
S(\s)) b(\s) e^{iS/\d}\\ - \ba_0(x,\n S(x))f_{\d,\e} +\ba_0(x,\n
S(x))f_{\d,\e} + o(1),
\end{multline*}
as $\d \ra 0$ for each fixed $\e >0$. So,
$$
\bL f_{\d,\e} - z f_{\d,\e} =\ln|\l(\s)| p^{-1}(\s) f_{\d,\e} +
o(1).
$$
By \eqref{e:eigval}, the logarithm is arbitrarily small, as $\e \ra
0$. Thus, letting $\d \ra 0$ first, then letting $\e \ra 0$
completes the proof in the periodic case.

{\sc Case 3: stagnation point.}

In the case of stagnation point we have $u(x_0) = 0$ and still $p(x)
<P$ for all $x \in U_{x_0}$.

First of all, we single out a simple situation when there is an open
neighborhood of $x_0$, denoted $V_{x_0}$, consisting entirely of
stagnation points of the flow $\f$. In this case we let $S$ be any
function such that $\n S(x_0) = \xi_0$. For $\e >0$ pick a function
$h_\e(x)$ of unit $L^2$-norm supported in $\{|x-x_0| <\e\}$ so that
for sufficiently small $\e$ the support of $h_\e$ is concentrated
inside $V_{x_0}$.

Since $\{\bB_t(x_0,\xi_0)b_0\}_{t \in \R}$ is bounded, the matrix
$\ba_0(x_0,\xi_0)$ has a purely imaginary eigenvalue $i\a$. Let
$v_0$ be the corresponding eigenvector. We set
$$
f_{\d,\e} = \bP[v_0 h_\e e^{iS/\d}].
$$
Then
\begin{align*}
\bL f_{\d,\e} - i\a f_{\d,\e} &= \ba_0(x,\n S(x))v_0 h_\e(x)
e^{iS(x)/\d} - i\a v_0 h_\e(x) e^{iS(x)/\d} + o(1) \\
& = (\ba_0(x,\n S(x))- i\a) v_0 h_\e(x) e^{iS(x)/\d} + o(1).
\end{align*}
As before, we let $\d \ra 0$ first, and then $\e \ra 0$.

Now, suppose that every neighborhood of $x_0$ contains a
non-stagnant point. Since the periods in a vicinity of $x_0$ are
bounded, $x_0$ is Lyapunov stable. This implies that for every
$\e>0$ there is a proper orbit $\mathcal{O}_\e$ contained entirely
in $\{|x - x_0| <\e\}$. By the density and invariance of $A$ we can
ensure the identity \eqref{E:star} on the orbit $\mathcal{O}_\e$.
Let $p_\e$ denote the prime period of $\mathcal{O}_\e$.

Since the orbit $\mathcal{O}_\e$ is contained in a small
neighborhood of $x_0$, we have
$$
\int_{\mathcal{O}_\e} u(x)\cdot \xi_0\ |dx| = 0.
$$
Hence, there is a point $x_\e \in \mathcal{O}_\e$ such that $u(x_\e)
\cdot \xi_0 = 0$.

Next, for each $\e>0$ we find a natural $N_\e$ such that
\begin{equation}\label{e:periods}
    P/3 < N_\e p_\e < P.
\end{equation}
As before, the matrix $\bB_1(x_0,\xi_0)$ has an eigenvalue $\l_0$
with $|\l_0| = 1$. So, by perturbation, the matrix $\bB_{N_\e
p_\e}(x_\e,\xi_0) = \bB_{p_\e}^{N_\e }(x_\e,\xi_0)$ has an
eigenvalue $\l_\e$ such that
\begin{equation}\label{e:eigenperiods}
    |\l_\e - \l_0^{N_\e p_\e}| = o(1), \text{ as } \e \ra 0.
\end{equation}
Observe that $\l_\e^{1/N_\e}$ is an eigenvalue of
$\bB_{p_\e}(x_\e,\xi_0)$. So, using the result of the previous
periodic case, we find
$$
p_\e^{-1} \ln(\l_\e^{1/N_\e}) \in \s(\bL).
$$
On the other hand,
$$
p_\e^{-1} \ln(\l_\e^{1/N_\e}) = (N_\e p_\e)^{-1} (\ln|\l_\e| + i
\arg \l_\e).
$$
By \eqref{e:periods} and \eqref{e:eigenperiods}, this sequence of
spectral points is bounded and at the same time the real parts
vanish as $\e \ra 0$. So, there is a subsequence converging to a
purely imaginary point.

Since the approximate eigenfunctions used in the proof are
weakly-null, the found point lies in the essential spectrum of the
generator.

This completes the proof of \eqref{e:sigmaperpincl}. \qed
\end{proof}

\section{Spectrum in Sobolev spaces $\sobcon$ for large
$|m|$}\label{S:sobolev}

In this section we describe the results concerning structure of the
essential spectrum over Sobolev spaces of sufficiently large
smoothness in the case when the basic flow has a nonzero Lyapunov
exponent.

First, we seek sufficient conditions for the spectrum $\Sigma_m$ to
be connected, i.e.
\begin{equation}\label{sigmaconn}
    \Sigma_m = [\mminm,\mmaxm].
\end{equation}
By \thm{T:sobdesc}, from \eqref{sigmaconn} we immediately obtain the
identity
\begin{equation}\label{specconn}
    |\ess{\bG_t}| = \exp\{t [\mminm,\mmaxm] \}.
\end{equation}
One trivial condition that guarantees \eqref{sigmaconn} follows from
Sacker and Sell's theorem stated in Section \ref{ss:cocycle}.
Namely, the dimension of the vector bundle $\frbundle$ is one.

Second, we identify certain margins of the spectrum, $[\mminm,s]
\cup [S,\mmaxm]$, which will be proved to satisfy the aperiodicity
assumption (ii) of \thm{T:rotinv}. A condition on $m$ will be found
that insures that these margins are nonempty, and the same condition
will imply \eqref{sigmaconn}. Hence, from \thm{T:rotinv} and
\eqref{specconn} we conclude that $\ess{\bG_t}$ contains solid
spectral rings and has no circular gaps. A generic configuration of
such spectrum is indicated in Figure \ref{F:semi}.

Third, we establish similar results for $\ess{\bL}$, and under the
same condition on $m$ we show that $\Sigma_m^\perp = \Sigma_m$. A
generic spectrum in this case is shown in Figure \ref{F:gen}.
Consequently, we obtain a variant of the spectral mapping property
for the group $\bG$, which in turn implies the identity between the
exponential type of $\bG$ and the spectral bound of $\bL$.

Before we state our results, let us introduce relevant notation.

Let $\lmin$ and $\lmax$ denote the end-points of the dynamical
spectrum of the $\xi$-cocycle $\p \f_t^{-\top}$. As we noted
earlier, this cocycle is the fundamental matrix solution of the
$\xi$-equation \eqref{BASxi}. Since $\p \f_t^{-\top}$ is the inverse
dual to the Jacobi cocycle $\p \f_t$ (see Section \ref{ss:cocycle}),
the end-points of the latter are $\llmin = - \lmax$, $\llmax =
-\lmin$. In view of the incompressibility assumption $\det(\p \f_t)
= 1$, the conditions $\lmax>0$ and $\llmax
>0$ are equivalent. If either of them holds we say that $\f$ has
{\bf exponential stretching of trajectories}. From now on we only
use the exponents $\lmax$ and $\lmin$.

Let us introduce the following constants
\begin{equation}\label{e:Ss}
s = \sup_{k \in \R}\{ \mu^k_{\mathrm{min}}\},\quad S = \inf_{k \in
\R}\{ \mu^k_{\mathrm{max}} \}.
\end{equation}
As we will see, any point of $\Sigma_m$ that lies outside the
interval $[s,S]$ (if such a point exists) satisfies the aperiodicity
assumption of \thm{T:rotinv}.

Our main result is stated in the following theorem.

\begin{theorem}\label{T:sobolev} Let $\bG$ be the $C_0$-group generated by equation
\eqref{pde} on $\sobcon$, $m\in \R$. Assume that $\lmax>0$ and $|m|
> \frac{\mmax - \mmin}{\lmax - \lmin}$. Then the following holds:
\begin{itemize}
  \item[1)] identities \eqref{sigmaconn} and \eqref{specconn} ;
  \item[2)] $\mminm < s$ and $S < \mmaxm$ ;
  \item[3)] $\T \cdot \exp \left\{ t [\mminm,s] \cup [S, \mmaxm] \right\}
    \ss \ess{\bG_t}$ ;
\end{itemize}
\end{theorem}

Thus, a generic spectral picture is reminiscent of the bicycle wheel
as shown in Figure \ref{F:semi}. The crucial features of the
spectrum in this case are two solid inner and outer rings, and no
circular spectral gap.

\begin{figure}
\centering
\psset{xunit=1mm,yunit=1mm,runit=1mm}
\begin{pspicture}(0,0)(120.00,65.00)
\pscircle[linewidth=0.15,linecolor=black,fillcolor=black,fillstyle=hlines,hatchwidth=0.28,hatchsep=1.42,hatchangle=45.00,hatchcolor=black](60.00,30.21){27.50}
\newrgbcolor{userFillColour}{1.00 1.00 1.00}
\pscircle[linewidth=0.15,linecolor=black,fillcolor=userFillColour,fillstyle=solid](60.00,30.21){21.12}
\psellipse[linewidth=0.15,linecolor=black,fillcolor=black,fillstyle=hlines,hatchwidth=0.28,hatchsep=1.42,hatchangle=45.00,hatchcolor=black](60.00,30.21)(11.79,11.79)
\newrgbcolor{userFillColour}{1.00 1.00 1.00}
\pscircle[linewidth=0.15,linecolor=black,fillcolor=userFillColour,fillstyle=solid](60.00,30.21){5.89}
\psbezier[linewidth=0.15,linecolor=black]{-}(68.49,38.62)(70.00,41.65)(71.10,42.76)(71.77,41.97)
\psbezier[linewidth=0.15,linecolor=black]{-}(71.77,41.97)(72.44,41.18)(73.50,42.23)(74.93,45.14)
\psbezier[linewidth=0.15,linecolor=black]{-}(45.25,15.21)(46.76,18.24)(47.86,19.35)(48.53,18.56)
\psbezier[linewidth=0.15,linecolor=black]{-}(48.53,18.56)(49.20,17.77)(50.26,18.82)(51.69,21.73)
\psbezier[linewidth=0.15,linecolor=black]{-}(38.93,30.88)(41.49,32.58)(43.11,32.59)(43.79,30.90)
\psbezier[linewidth=0.15,linecolor=black]{-}(43.79,30.90)(44.48,29.21)(45.95,29.18)(48.21,30.79)
\psbezier[linewidth=0.15,linecolor=black]{-}(46.12,46.09)(47.17,43.12)(48.26,42.00)(49.38,42.74)
\psbezier[linewidth=0.15,linecolor=black]{-}(49.38,42.74)(50.51,43.49)(51.60,42.37)(52.65,39.40)
\psbezier[linewidth=0.15,linecolor=black]{-}(60.25,51.35)(61.71,48.72)(61.82,47.13)(60.59,46.57)
\psbezier[linewidth=0.15,linecolor=black]{-}(60.59,46.57)(59.35,46.01)(59.29,44.46)(60.42,41.91)
\psbezier[linewidth=0.15,linecolor=black]{-}(71.79,28.46)(74.63,29.46)(76.16,29.25)(76.38,27.84)
\psbezier[linewidth=0.15,linecolor=black]{-}(76.38,27.84)(76.61,26.43)(78.06,26.07)(80.73,26.78)
\psbezier[linewidth=0.15,linecolor=black]{-}(68.53,22.19)(69.63,19.10)(70.68,17.81)(71.67,18.33)
\psbezier[linewidth=0.15,linecolor=black]{-}(71.67,18.33)(72.66,18.84)(73.73,17.87)(74.88,15.42)
\psbezier[linewidth=0.15,linecolor=black]{-}(60.16,18.42)(61.38,16.13)(61.31,14.58)(59.94,13.78)
\psbezier[linewidth=0.15,linecolor=black]{-}(59.94,13.78)(58.57,12.99)(58.62,11.44)(60.09,9.15)
\psline[linewidth=0.15,linecolor=black,linestyle=dotted,dotsep=1.00]{->}(20.00,30.21)(100.00,30.21)
\psline[linewidth=0.15,linecolor=black,linestyle=dotted,dotsep=1.00]{->}(60.00,0.21)(60.00,60.21)
\rput(92.5,32.00){{\tiny $e^{\mmaxm t}$}}%
\rput(78.5,32.00){{\tiny $e^{S t}$}} \rput(74,32.00){{\tiny $e^{ s
t}$}} \rput(61,32.00){{\tiny $e^{ \mminm t}$}}
\end{pspicture}

\caption{"Bicycle wheel" structure of the essential spectrum of
${\bG_t}$ over $\sobcon$, under the assumptions of \thm{T:sobolev}.}
\label{F:semi}

\end{figure}

Similar description can be given to the spectrum of the generator.

\begin{theorem}\label{T:sobolevgen}
Under the assumption of \thm{T:sobolev} the following holds:
\begin{itemize}
\item[1)] $\Sigma_m^\perp =  \Sigma_m = [\mminm,\mmaxm]$ ;
\item[2)] $[\mminm,s]\cup[S,\mmaxm] + i\R \ss \ess{\bL}$ ;
\item[3)] $\re \ess{\bL} = [\mminm,\mmaxm]$.
\end{itemize}
\end{theorem}

A generic spectral picture in this case is shown in Figure
\ref{F:gen}. Results of \cite{Lif-elliptic} show that the ladder
structure is possible for elliptic flows, even though those have no
exponential stretching.

\begin{figure}

\centering




\psset{xunit=1mm,yunit=1mm,runit=1mm}

\begin{pspicture}(0,0)(120.00,70.00)

\psline[linewidth=0.15,linecolor=black,fillcolor=black,fillstyle=hlines,hatchwidth=0.28,hatchsep=1.42,hatchangle=45.00,hatchcolor=black]{-}(40.00,0.00)(40.00,60.00)(40.00,65.00)(42.50,62.50)(45.00,65.00)(47.50,62.50)(50.00,65.00)(50.00,0.00)(47.50,2.50)(45.00,0.00)(42.50,2.50)(40.00,0.00)(40.00,0.00)(40.00,0.00)
\psline[linewidth=0.15,linecolor=black,fillcolor=black,fillstyle=hlines,hatchwidth=0.28,hatchsep=1.42,hatchangle=45.00,hatchcolor=black]{-}(70.00,0.00)(70.00,60.00)(70.00,65.00)(72.50,62.50)(75.00,65.00)(77.50,62.50)(80.00,65.00)(80.00,0.00)(77.50,2.50)(75.00,0.00)(72.50,2.50)(70.00,0.00)(70.00,0.00)(70.00,0.00)
\psbezier[linewidth=0.15,linecolor=black]{-}(50.00,32.50)(57.50,25.00)(65.00,42.50)(70.00,32.50)
\psbezier[linewidth=0.15,linecolor=black]{-}(50.00,50.00)(57.50,60.00)(62.50,40.00)(70.00,50.00)
\psbezier[linewidth=0.15,linecolor=black]{-}(50.00,15.00)(60.00,22.50)(60.00,7.50)(70.00,15.00)
\psline[linewidth=0.15,linecolor=black,linestyle=dotted,dotsep=1.00]{<-}(60.00,70.00)(60.00,-5.00)
\psline[linewidth=0.15,linecolor=black,linestyle=dotted,dotsep=1.00]{->}(30.00,30.00)(90.00,30.00)
\rput(85,27.50){{\small $\mmaxm$}} \rput(67.50,27.50){{\small $S$}}
\rput(52.50,27.50){{\small $s$}} \rput(35,27.50){{\small $\mminm$}}

\end{pspicture}

\caption{"Ladder" structure of the essential spectrum of $\bL$ over
$\sobcon$ under the assumptions of \thm{T:spgen}.}\label{F:gen}

\end{figure}

Combining the above theorems with \eqref{discexp} we obtain the
following spectral mapping property.

\begin{theorem}[Annular Hull Theorem]\label{T:aht}
Under the assumptions of \thm{T:sobolev} one has
\begin{equation}\label{e:aht}
 \T \cdot \s(\bG_t) = \exp\{ t \s(\bL) + i\R \}.
\end{equation}
\end{theorem}
As a consequence, we obtain the identity between the exponential
type of the semigroup, and the spectral bound of the generator:
\begin{equation}\label{eq:w=s}
    \w(\bG) = s(\bL).
\end{equation}

The proofs of Theorems \ref{T:sobolev} and \ref{T:sobolevgen} given
in the next section will come out as a result of systematic study of
the dynamical spectrum $\Sigma_m$.

Let us consider now one particular case when $\Sigma = \{0\}$, i.e.
$\mmax = \mmin = 0$. So, the amplitude equation has no exponentially
growing or decaying solutions. Examples from our list that trivially
satisfy this condition are the 2D Euler in vorticity formulation,
simple transport, SQG, and CHM equations. In this case the
hypothesis of \thm{T:sobolev} is satisfied for all $m \neq 0$
provided $\lmax >0$. Thus, $S \leq 0 \leq s$, and from Theorems
\ref{T:sobolev} and \ref{T:sobolevgen} we obtain the following
corollary.
\begin{corollary}\label{sobcorol}
Suppose that $\lmax >0$ and $\Sigma = \{0\}$. Then for any $m \neq
0$ one has the identities
\begin{align}
\ess{\bG_t} &= \T \cdot \exp\{ t m [\lmin,\lmax] \}, \label{sigma01} \\
\ess{\bL} &= i \R + m [\lmin,\lmax] \label{sigma02}
\end{align}
over the space $\sobcon$.
\end{corollary}

In particular, the full spectral mapping theorem holds. Moreover, if
$n=2$, then from $\det \p \f_t = 1$ we have $\lmin = - \lmax$. So,
the essential spectrum of the 2D Euler and SQG equations is a solid
band (annulus) symmetric with respect to the imaginary axis. This
result was obtained previously by Latushkin, Friedlander and the
author in \cite{SL2003b,SL2003a,FS2004} via an explicit construction
of approximate eigenfunctions for each point in the band.

In the case $m=0$ the identities \eqref{sigma01}, \eqref{sigma02}
become inclusions $\subseteq$ due to \thm{T:sobdesc}. These again
turn into identities provided $u$ has arbitrarily long trajectories
\cite{SL2003a}.

\section{Dynamical spectrum of the $b\xi^m$-cocycle} \label{s:sss}

In this section we present the proofs of Theorems \ref{T:sobolev}
and \ref{T:sobolevgen}.

We introduce a scalar cocycle $\bX^m$, the $\xi^m$-component of the
$b\xi^m$-cocycle, by the rule
\begin{equation}\label{Xcocycle}
    \bX^m_t(x,\xi) = \left|\p \f_t^{-\top}(x) \overline{\xi}
    \right|^m.
\end{equation}
Notice that $\bX^m$ is one-dimensional and is defined on the trivial
scalar bundle over $\O^n$ (or $\K^n$). Hence, by Sacker and Sell's
theorem its spectrum consists of a single interval given by
\begin{equation}\label{ximspec}
    \Sigma_{\bX^m} = m [\lmin,\lmax].
\end{equation}

We notice that the $b\xi^m$-cocycle is isomorphic to the tensor
product of $\bX^m$ and $\bB$. Thus, from the results of
\cite{ShvCocycle} we obtain the following proposition.
\begin{proposition}
The following inclusion holds:
\begin{equation}\label{e:inclbxi1}
\Sigma_m \ss \Sigma + m [\lmin,\lmax].
\end{equation}
\end{proposition}

Certain estimates on the end-points of $\Sigma_m$ follow trivially
by definition or from \eqref{e:inclbxi1}. Let us denote
$$
\begin{aligned}
A_m &=  \mmin + m \lmin \\
B_m &=  \mmax + m \lmin
\end{aligned}
\text{\quad ; \quad}
\begin{aligned}
C_m &=  \mmin + m \lmax  \\
D_m &=  \mmax + m \lmax
\end{aligned}
$$
for positive $m$, and
$$
\begin{aligned}
A_m &=  \mmin + m \lmax  \\
B_m &=  \mmax + m \lmax
\end{aligned}
\text{\quad ; \quad}
\begin{aligned}
C_m &=  \mmin + m \lmin \\
D_m &=  \mmax + m \lmin
\end{aligned}
$$
for negative $m$.

\begin{lemma}\label{L:estimates}
The following estimates hold for all $m \in \R$:
\begin{align}
A_m &\leq \mminm \leq \min\{B_m,C_m\}, \label{E:estmin}\\
\max\{B_m,C_m\} &\leq \mmaxm \leq D_m. \label{E:estmax}
\end{align}
\end{lemma}

It is clear from these estimates that exponential stretching causes
expansion of the spectrum $\Sigma_m$, as $m \ra \infty$. On the
other hand, if $\lmax = 0$, it follows from \eqref{e:inclbxi1}  that
$\Sigma_m \ss \Sigma$. Likewise, since $\bB = \bX^{-m} \BX$, we have
$\Sigma \ss \Sigma_m$.  So, we have proved the following lemma.
\begin{lemma}\label{L:change}
The identity $\Sigma = \Sigma_m$ holds for all $m \in \R$ if and
only if the flow $\f$ has no exponential stretching of trajectories.
\end{lemma}

Generally, $\Sigma_m$ may have gaps. We can estimate the location of
a possible gap in $\Sigma_m$ using the Lyapunov exponents of the
flow $\f$.

\begin{proposition}\label{p:gap}
If $m$ is positive, then
\begin{equation}\label{gap1}
    [\mminm,\mmaxm] \backslash \Sigma_m \ss [\mmin + m \lmax\,,\, \mmax + m \lmin].
\end{equation}
If $m$ is negative, then
\begin{equation}\label{gap2}
    [\mminm,\mmaxm] \backslash \Sigma_m \ss [\mmin + m \lmin\, ,\, \mmax + m \lmax].
\end{equation}
\end{proposition}

\begin{proof}
Let $\mu \in [\mminm,\mmaxm]$ belong to the resolvent set of the
$b\xi^m$-cocycle. Then there is an exponential dichotomy of the
rescaled cocycle over $\K^n$. Let $\P$, $\e$ and $M$ be as in the
definition (see Section \ref{ss:cocycle}). Since the projector $\P$
is non-trivial for all $(x,\xi) \in \K^n$, there
exist $b_1,\,b_2 \in F(\xi)$ of unit norm such that%
\begin{align}
b_1  \in \Rg \P(x,\xi),\quad b_2  \in \Ker \P(x,\xi).
\end{align}
Thus, we have
\begin{subequations}\label{en:dich}
\begin{align}
|\bB_t(x,\xi)b_1| \cdot | \p \f^{-\top}_t(x)\xi|^m &\leq M
e^{t(\mu-\e)}, \label{en:dich1}\\
|\bB_t(x,\xi)b_2| \cdot | \p \f^{-\top}_t(x)\xi|^m &\geq M^{-1}
e^{t(\mu+\e)}, \label{en:dich2}
\end{align}
\end{subequations}
for all $t \geq 0$.

Let $\l$ be any end-point of $m[\lmin,\lmax]$. It is an exact
Lyapunov exponent of $\bX^m$ by \cite{JPS87}. So, there is a point
$(x,\xi) \in \K^n$ such that
\begin{equation}\label{e:asymlyap}
e^{(\l-\e)t} \leq | \p \f^{-\top}_t(x)\xi |^m \leq e^{(\l+\e)t},
\end{equation}
for sufficiently large $t$. Combining \eqref{e:asymlyap} with the
inequalities in \eqref{en:dich}, we obtain
\begin{subequations}\label{en:conseq}
\begin{align}
|\bB_t(x,\xi) b_1| &\leq M e^{t(\mu-\l)}, \\
|\bB_t(x,\xi) b_2| &\geq M^{-1} e^{t(\mu-\l)}.
\end{align}
\end{subequations}
Considering that the exponential type of $|\bB_t(x,\xi) b_1|$ is not
less than $\mmin$, while the exponential type of $|\bB_t(x,\xi)
b_2|$ does not exceed $\mmax$, \eqref{en:conseq} imply $\l +\mmin
\leq \mu \leq \l + \mmax$. This proves the proposition.\qed
\end{proof}

We can see from \eqref{gap1} and \eqref{gap2} that for large values
of $|m|$ exponential stretching closes gaps in $\Sigma_m$.
Precisely, $|m|$ has to be such that the end-points of the intervals
on the left-hand sides of \eqref{gap1} and \eqref{gap2} meet. Thus,
we obtain the following corollary, which in turn proves part 1) of
\thm{T:sobolev}.

\begin{corollary}\label{c:connected}
If $\lmax>0$ and $|m| \geq \frac{\mmax - \mmin}{\lmax - \lmin}$,
then $\Sigma_m$ is connected.
\end{corollary}

\subsection{The marginal spectrum}\label{S:margin}

Let us recall the following constants introduced in Section
\ref{S:sobolev}:
\begin{equation}\label{e:Ss2}
s = \sup_{k \in \R}\{ \mu^k_{\mathrm{min}}\},\quad S = \inf_{k \in
\R}\{ \mu^k_{\mathrm{max}} \}.
\end{equation}
We define {\bf marginal spectrum} of the $b\xi^m$-cocycle as the set
\begin{equation}\label{d:margin}
    \Mg_m = \text{closure of } \Sigma_m \cap \left[ (-\infty, s) \cup (S, +\infty)
    \right].
\end{equation}
We show that to any point of $\Mg_m$ there corresponds a \Mp\ in the
sense of \thm{T:sp}, that is surrounded by non-periodic
exponentially stretched orbits. But first, we find conditions which
guarantee that $\Mg_m$ is non-empty, i.e. $\mmaxm
> S$ and $\mminm<s$.

According to Lemma \ref{L:change}, a non-empty marginal spectrum is
possible only if the flow $\f$ has exponential stretching. In view
of estimates \eqref{E:estmin} and \eqref{E:estmax}, it suffices to
have $B_m<\mmin $ and $C_m
> \mmax$. We see that both inequalities are satisfied under the assumption \thm{T:sobolev},
which in view of Corollary \ref{c:connected} also implies
connectedness of the two sides of $\Mg_m$. Thus, we have obtained
the following lemma.

\begin{lemma}\label{L:nonempty}
Assume that $\lmax>0$ and $|m| > \frac{\mmax - \mmin}{\lmax -
\lmin}$. Then $\Mg_m$ is non-empty and described by the following
identity:
\begin{equation}\label{e:descmargin}
    \Mg_m = [\mminm, s] \cup [S, \mmaxm].
\end{equation}
\end{lemma}

We show now that to every point of the marginal spectrum there
corresponds a \Mp\ from $\O^n_0$ that satisfies the aperiodicity
condition (ii) of \thm{T:rotinv}.

\begin{proposition}\label{p:margin} For all $m \in \R$ we have
\begin{equation}\label{e:marginident}
   \Mg_m = \Mg_m \cap \Sigma_m^\perp.
\end{equation}
Furthermore, for any element of $\Mg_m$ there exists a \Mp\
$(x_0,\xi_0) \in \O^n_0$, corresponding either to the rescaled
$b\xi^m$-cocycle or to its dual, such that every neighborhood of
$x_0$ intersects a non-periodic orbit.
\end{proposition}
\begin{proof}
We present the proof for the right margin only, since the assertion
for the left margin follows by passing to the inverse cocycle.

We write $f(t) \lesssim g(t)$ to signify that the exponential type
of $f(t)$ is less than the exponential type of $g(t)$.

If $\mmaxm \leq S$, then the identity \eqref{e:marginident} is
trivial. So, let $\mmaxm > \mmaxk$ for some $k \in \R$, and we fix
any $\l \in \Sigma_m$ with $\l > \mmaxk$.

We proceed in several steps considering all possible cases.

Suppose $m>k$. According to \thm{T:sp} there is a Ma\~{n}e sequence
either for the $b\xi^m$-cocycle or for its dual.

Let us assume the former. Then by \eqref{e:mane1a} there is a
sequence $\{(x_n,\xi_n),b_n\}_{n=1}^\infty$, with $b_n \in
F(\xi_n)$, such that
$$
|b(n)||\xi(n)|^m \gtrsim e^{n \l}.
$$
Here we denote
\begin{align*}
b(n) &= \bB_n(x_n,\xi_n)b_n, \\
\xi(n) & = \p \f_n^{-\top}(x_n)\xi_n.
\end{align*}

At the same time, by the definition of $\mmaxk$, one has
$$
|b(n)||\xi(n)|^m = |b(n)||\xi(n)|^k |\xi(n)|^{m-k}\lesssim
e^{n\mmaxk} |\xi(n)|^{m-k}.
$$
This shows that $\xi(n)$ is growing exponentially as $n \ra \infty$.

Let $(x_0,\xi_0)$, $|\xi_0| = 1$, be the corresponding Ma\~{n}e
point constructed from the sequence
$\{(x_n,\overline{\xi}_n)\}_{n=1}^\infty \ss \K^n$ as in Section
\ref{ss:cocycle}. Thus, by construction,
\begin{align*}
x_0 &= \lim_{n \ra \infty} \f_n(x_n),\\
\xi_0 &= \lim_{n \ra \infty}\frac{\xi(n)}{|\xi(n)|}.
\end{align*}
It follows that
\begin{equation*}
u(x_0)\cdot \xi_0 = \lim_{n \ra \infty} u(\f_{n}(x_n)) \cdot
\frac{\xi(n)}{|\xi(n)|} = \lim_{n \ra \infty} u(x_n) \cdot
\frac{\xi_n}{|\xi(n)|} = 0.
\end{equation*}
Thus, $(x_0,\xi_0)$ belongs to $\O^n_0$, and hence, $\l \in
\Mg_m^\perp$.

Suppose now there is a Ma\~{n}e sequence for the dual cocycle, while
still $m>k$. Then there exists a Ma\~{n}e point $(x_0,\xi_0) \in
\K^n$ and  Ma\~{n}e vector $b_0 \in F(\xi_0)$. It is straightforward
to check the the dual cocycle is given by
\begin{equation}\label{bxidual}
\left|\p\f_{-t}^{-\top}(x)\overline{\xi}\right|^{-m}\bB_{-t}^{-*}(x,\xi).
\end{equation}
Thus, we have
$$
|\bB_{-t}^{-*}(x_0,\xi_0)b_0| \cdot |\p
\f_{-t}^{-\top}(x_0)\xi_0|^{-m} \leq C e^{\l t}, \quad t \in \R.
$$
We can see that the same point is Ma\~{n}e for the inverse dual
cocycle rescaled by $-\l$, so that
$$
|\bB_{t}^{-*}(x_0,\xi_0)b_0| \cdot |\p
\f_{t}^{-\top}(x_0)\xi_0|^{-m} \leq C e^{-\l t}, \quad t \in \R.
$$

Let us reconstruct the \Ms\ from $(x_0,\xi_0,b_0)$ as in Section
\ref{ss:cocycle}. We obtain a sequence
$\{(x_n,\xi_n),b_n\}_{n=1}^\infty$, with $b_n \in F(\xi_n)$, such
that, in particular,
\begin{equation}\label{E:A}
|b^*(n)||\xi(n)|^{-m} \lesssim e^{-\l n},
\end{equation}
where $\xi(n)$ as before, and
$$
b^*(n) =\bB_{n}^{-*}(x_n,\xi_n)b_n.
$$

Recall that the inverse dual cocycle has the opposite spectrum.
Thus, in addition to \eqref{E:A}, we obtain
\begin{equation}\label{E:B}
|b^*(n)||\xi(n)|^{-m} \gtrsim e^{-n \mmaxk} |\xi(n)|^{k-m}.
\end{equation}
Combining \eqref{E:A} and \eqref{E:B} we see again that $\xi(n)$ is
exponentially increasing.

We find a point $(x'_0,\xi'_0) \in \O_0^n$ constructed from this
sequence, which is a Ma\~{n}e point for the inverse dual cocycle
rescaled by $-\l$. As before, this same point is Ma\~{n}e for the
dual cocycle rescaled by $\l$. So, $\l$ belongs to the dynamical
spectrum of the dual cocycle, and hence, to the spectrum of the
original $b\xi^m$-cocycle on $\O_0^n$. Thus, $\l \in
\Sigma_m^\perp$.

Let us now consider the case when $m<k$. If a Ma\~{n}e sequence
exists for the $b\xi^m$-cocycle itself, then as in the previous
paragraph we can find a Ma\~{n}e sequence for the inverse cocycle
corresponding to $-\l$. This implies
$$
e^{-n \mmaxk} |\xi(-n)|^{m-k} \lesssim |b(-n)||\xi(-n)|^m \lesssim
e^{-n \l}.
$$
So, $|\xi(-n)|$ is exponentially increasing and we finish the proof
as before.

Finally, if the dual to the $b\xi^m$-cocycle has a Ma\~{n}e
sequence, then we obtain
$$
e^{n \mmaxk} |\xi(-n)|^{k-m} \gtrsim |b^*(-n)||\xi(-n)|^{-m} \gtrsim
e^{n \l}.
$$
Again, $|\xi(-n)|$ is increasing exponentially.

We have considered all possible cases. This finishes the proof of
\eqref{e:marginident}.

Let us notice that in each of the above cases we have obtained a
Ma\~{n}e point $(x_0,\xi_0) \in \O^n_0$ such that $\xi(t,x_0,\xi_0)$
has a non-trivial exponential type either in the forward or in the
backward direction. If the $\f$-orbit through $x_0$ is not periodic,
then the "furthermore'' statement is trivial. Otherwise, we use the
Stable Manifold Theorem \cite{Perko} to find stable and unstable
manifolds in a neighborhood of the orbit. Either of the manifolds
consists of non-periodic orbits.\qed
\end{proof}

\begin{corollary}\label{c:perp}
If $\lmax>0$ and $|m| > \frac{\mmax - \mmin}{\lmax - \lmin}$, then
\begin{equation}\label{e:perp=sigma}
\Sigma_m^\perp = \Sigma_m = [\mminm,\mmaxm]\end{equation}
\end{corollary}
\begin{proof}
The end-points of $\Sigma_m$ and $\Sigma_m^\perp$ coincide due to
Proposition \ref{p:margin}. On the other hand, the analogue of
Proposition \ref{p:gap} on $\O^n_0$ is straightforward. It only
suffices to notice that if $\lmax>0$, then the point $(x,\xi)$ in
formula \eqref{e:asymlyap} must belong to $\O^n_0$ by conservation
of the Hamiltonian $u(x) \cdot \xi$.

As in Corollary \ref{c:connected} this shows that $\Sigma_m^\perp$
is connected, and hence, \eqref{e:perp=sigma} holds. \qed
\end{proof}

With the help of \thm{T:rotinv} and \thm{T:spgen} the results of
this section prove Theorems \ref{T:sobolev} and \ref{T:sobolevgen}
completely.

\def\cprime{$'$} \def\cprime{$'$} \def\cprime{$'$} \def\cprime{$'$}
  \def\cprime{$'$} \def\cprime{$'$} \def\cprime{$'$} \def\cprime{$'$}
\providecommand{\bysame}{\leavevmode\hbox
to3em{\hrulefill}\thinspace}
\providecommand{\MR}{\relax\ifhmode\unskip\space\fi MR }
\providecommand{\MRhref}[2]{%
  \href{http://www.ams.org/mathscinet-getitem?mr=#1}{#2}
} \providecommand{\href}[2]{#2}

\end{document}